\documentclass[10pt,aps,prx,twocolumn,notitlepage,showpacs,superscriptaddress,longbibliography]{revtex4-2}

\usepackage{times}
\usepackage{amssymb,amsmath}
\usepackage{bm}
\usepackage{graphicx}
\usepackage{xcolor}

\usepackage[urlcolor=blue,colorlinks=true,citecolor=blue,linkcolor=blue,pdfstartview={FitH},bookmarks=false]{hyperref}
\newcommand{\kap}[1]{\boldsymbol{\kappa}}

\begin{document}
	
\title{Instabilities of Fermi Liquids with Arbitrary Forward Scattering: Exact Approach}

\date{\today}
\author{Dmitry Miserev$^{\ast}$ }\affiliation{Department of Physics, University of Basel, Klingelbergstrasse 82, 4056 Basel, Switzerland}

\author{Joel Hutchinson}\affiliation{Department of Physics, University of Basel, Klingelbergstrasse 82, 4056 Basel, Switzerland}

\author{Herbert Schoeller}\affiliation{Institut f\"{u}r Theorie der Statistischen Physik, RWTH Aachen University and JARA -- Fundamentals of Future Information Technology,
	52056 Aachen, Germany}

\author{Jelena Klinovaja}\affiliation{Department of Physics, University of Basel, Klingelbergstrasse 82, 4056 Basel, Switzerland}

\author{Daniel Loss} \affiliation{Department of Physics, University of Basel, Klingelbergstrasse 82, 4056 Basel, Switzerland}
%\affiliation{RIKEN Center for Quantum Computing (RQC), 2-1 Hirosawa, Wako, Saitama 351-0198, Japan}
\affiliation{Physics Department, King Fahd University of Petroleum and Minerals, 31261, Dhahran, Saudi Arabia}
\affiliation{Quantum Center, KFUPM, Dhahran, Saudi Arabia}
\affiliation{RDIA Chair in Quantum Computing}

\begin{abstract}
	In this work, we consider $N$-fold degenerate $D$-dimensional electron gas with spherical Fermi surface and arbitrary forward-scattering density-density interaction transferring small momentum compared to the Fermi momentum $k_{\mathrm{F}}$. The dimensional reduction that is mathematically equivalent to the Haldane patch construction and similar multidimensional bosonization techniques, provides a natural map of two-point $D$-dimensional correlation functions (fermion Green function, susceptibilities etc.) onto effective one-dimensional (1D) correlators with the same diagrammatic structure, which can be evaluated exactly within a 1D bosonizable (Gaussian) theory. We then apply this formalism to evaluate the fermion Green function, pair and charge/flavor susceptibilities, as well as the composite correlation functions for the case of a finite-range interaction, where the interaction range $R_{\mathrm{s}} \gg 1/k_{\mathrm{F}}$ is large compared to the Fermi wavelength.
	First, we find that the single-particle spectral function remains Fermi-liquid-like which is fully consistent with the previous research.
	In contrast to the single-particle sector, the many-body channels are efficiently dressed by finite-range interactions, and this dressing is fully equivalent to the one-loop renormalization group (RG), which is also in line with previous multidimensional bosonization results. 
	Within the forward-scattering model, stable long-range order is not possible, and relevant susceptibilities demonstrate singular power-law scaling with temperature $T$ at $T \to 0$.
	However, the long-range order can be further stabilized by weak short-range interactions that are present in realistic physical systems.
	We also calculated composite susceptibilities, and found that the $2 N e$ pair susceptibility is more relevant than usual $2 e$ pair susceptibility provided the attractive coupling constant is large enough. 
	Similarly, $2 N k_{\mathrm{F}}$ density-wave susceptibilities are more relevant than usual $2 k_{\mathrm{F}}$ susceptibilities at large enough repulsive interactions.
	In particular, in case of the two-fold degenerate electron gas ($N = 2$), we predict $4 e$ superconductivity for large enough attractive interactions, and four-fermion $4 k_{\mathrm{F}}$ density waves for strong enough repulsive interactions.
\end{abstract}

\maketitle

\section{Introduction}
\label{sec:intro}

Interacting one-dimensional (1D) fermion metals at low temperatures realize strongly correlated Luttinger liquids, whose elementary excitations are gapless spin and charge modes; long-lived fermionic quasiparticles do not exist in this regime \cite{delftschoeller,giamarchi}. 
By contrast, in spatial dimensions greater than one ($D > 1$), interacting fermion metals behave as Fermi liquids with well-defined Landau quasiparticles characterized by the renormalized effective mass and finite, though parametrically large, lifetime $\propto E_{\mathrm{F}}/T^2 \gg 1/T$, where $E_{\mathrm{F}}$ is the Fermi energy, and $T \ll E_{\mathrm{F}}$ the temperature \cite{LandauFermiliquid,LandauFermiliquid2,AbrikosovKhalatnikov1959FermiLiquid,GiulianiQuinn1982Lifetime2DEG}. 
Despite this fundamental difference between 1D and higher-dimensional metals, $D$-dimensional correlation functions can still be mapped onto 1D correlators and evaluated within a 1D theory endowed with a suitably chosen effective interaction \cite{miserevDimensionalReductionLuttingerWard2023a}. 
This strategy of multidimensional bosonization --- originally known as the Haldane patch construction \cite{haldaneLuttingerTheoremBosonization2005} --- applies to small-angle scattering, also known as forward-scattering or finite-range interactions controlled by the small parameter $(k_{\mathrm{F}} R_{\mathrm{s}})^{-1} \ll 1$, where $k_{\mathrm{F}}$ is the Fermi momentum and $R_{\mathrm{s}}$ the interaction range. 
The Haldane patch construction yields the leading contributions in an expansion in $(k_{\mathrm{F}} R_{\mathrm{s}})^{-1}$.
Several mathematically equivalent formulations of the multidimensional bosonization exist, including the Ward-identity approach \cite{metznerFermiSystemsStrong1998} rooted in the Dzyaloshinskii-Larkin derivation of the 1D fermion Green function \cite{dzyaloshinskiilarkin}, the renormalization group (RG) \cite{Shankar1994RGFermiLiquid}, the Fermi-surface bosonization \cite{delacretazNonlinearBosonizationFermi2022}, the effective action approach \cite{kopietzBosonizationInteractingFermions2006,HoughtonKwonMarston2000MultidimensionalBosonization,BaresWen1993ElectronSpectralFunction,HoughtonKwonMarstonShankar1994CoulombBosonization,KwonHoughtonMarston1995TheoryFermionLiquids,castronetoBosonizationFermiLiquids1994,frohlichEffectiveGaugeField1995,efetovExactBosonizationInteracting2009}, and the dimensional reduction of the Luttinger–Ward functional \cite{miserevDimensionalReductionLuttingerWard2023a}. 
Here, we adopt the dimensional-reduction approach \cite{miserevDimensionalReductionLuttingerWard2023a}, which provides a general framework for evaluation of arbitrary $D$-dimensional correlation functions.
This approach was recently used to calculate charge, spin, and pair susceptibilities of a $D$-dimensional Fermi liquid with a finite-range interaction \cite{miserevMicroscopicMechanismPair2024}: various $2 k_{\mathrm{F}}$ density-wave orders, including an elusive pair-density-wave order, are strongly promoted by a finite-range interaction resulting in the power-law singularity of the corresponding susceptibility at $T \to 0$ when the interaction coupling constant exceeds the critical value.
In particular, a recently proposed non-BCS mechanism of high-temperature superconductivity (HTSC) driven by attractive finite-range interactions accounts for key features of HTSC, such as the dome-shaped dependence of the critical temperature $T_{\mathrm{c}}$ on doping, strongly suppressed isotope effect, and large $T_{\mathrm{c}} \sim 0.1 E_{\mathrm{F}}$ near the optimal doping \cite{miserev_high-temperature_2025}.

There are two major results of various multidimensional bosonization approaches: (i) dressed $D$-dimensional interaction at small momentum transfer $q \ll k_{\mathrm{F}}$ and small frequency $\omega \ll E_{\mathrm{F}}$ can be described by the random phase approximation (RPA) \cite{metznerFermiSystemsStrong1998,frohlichEffectiveGaugeField1995}, (ii) the fermion Green function demonstrates the Fermi-liquid behavior through a finite quasiparticle residue and a finite density of states at the Fermi level \cite{metznerFermiSystemsStrong1998,Shankar1994RGFermiLiquid}.
The multidimensional bosonization approach has been also applied to studies of the nematic quantum phase transition and non-Fermi-liquid states with singular interactions \cite{KwonHoughtonMarston1995TheoryFermionLiquids,khveshchenkoLowenergyPropertiesTwodimensional1993,altshulerLowenergyPropertiesFermions1994,chubukov_effect_2006}.

The single-particle Green function calculated within the leading order in $1/(k_{\mathrm{F}} R_{\mathrm{s}})$ demonstrates finite density of states at the Fermi surface and finite quasiparticle residue, thus reinforcing the Landau-Fermi liquid theory \cite{metznerFermiSystemsStrong1998,Shankar1994RGFermiLiquid}.
However, this does not mean that the Fermi-liquid instabilities instigated by finite-range interactions can be analyzed by means of the Landau-Fermi liquid theory.
In fact, it has been explicitly shown in Ref.~\cite{miserevMicroscopicMechanismPair2024} that, within the one-loop RG, various susceptibilities acquire non-perturbative power-law factors enhancing charge- and spin-density-wave susceptibilities for repulsive and the pair susceptibility for attractive interactions.
Within the leading order in $1/(k_{\mathrm{F}} R_{\mathrm{s}})$, all susceptibilities remain finite at any finite temperature $T > 0$, and diverge as $\propto 1/T^{|\gamma|}$ at $T \to 0$ and $\gamma < 0$ for the pair susceptibility, $\propto 1/T^{\gamma - \gamma_{\mathrm{c}}}$ for charge and spin density wave susceptibilities at $\gamma > \gamma_{\mathrm{c}}$, where $\gamma < 0$ ($\gamma > 0$) is the dimensionless coupling constant corresponding to attractive (repulsive) interaction.
Here, $\gamma_{\mathrm{c}} = (D - 1)/2$ is the critical coupling determining the onset of $D$-dimensional charge/spin density waves \cite{miserevMicroscopicMechanismPair2024}.
The long-range orders remain unstable within the leading $1/(k_{\mathrm{F}} R_{\mathrm{s}})$ expansion. 
However, a small residual short-range interaction can stabilize the long-range order resulting in a quantum phase transition.
This mechanism has been considered in detail for the pair susceptibility calculated in presence of a finite-range attractive interaction with small attractive short-range part \cite{miserev_high-temperature_2025}.
This theory, when applied to cuprates, correctly predicts main features of HTSCs such as the dome-shaped dependence of the critical temperature, $T_{\mathrm{c}}$, versus doping, weak isotope effect, and close-to-experimental values of the optimal doping and maximal $T_{\mathrm{c}}$ \cite{miserev_high-temperature_2025}.

In this work, we derive two-point correlation functions of interacting $D$-dimensional electron gas with $N$ fermion flavors (spin, valley, etc.) within the leading order in $1/(k_{\mathrm{F}} R_{\mathrm{s}})$. 
For simplicity, we assume a spherical Fermi surface with the Fermi velocity $v_{\mathrm{F}}$ that is the same at every point of the Fermi surface and for all $N$ fermion species.
The presented approach has a straightforward generalization for non-spherical Fermi surfaces with non-vanishing Gauss curvature and Fermi velocities depending on the position on the Fermi surface and the fermion flavor.
In the case of finite-range interactions, the single-particle Green function retains its Fermi-liquid form, which agrees with previous research, e.g. see Ref.~\cite{metznerFermiSystemsStrong1998}.
However, finite-range interactions result in the power-law enhancement of relevant susceptibilities (pairing susceptibilities for attractive and density-wave susceptibilities for repulsive interactions) which is fully equivalent to the one-loop RG calculated earlier in Ref.~\cite{miserevMicroscopicMechanismPair2024}.
This is also consistent with Ref.~\cite{Shankar1994RGFermiLiquid} where it has been shown that the coupling constant is not running in the leading order with respect to $1/(k_{\mathrm{F}} R_{\mathrm{s}})$.
In this work, we also calculate higher-order susceptibilities probing instabilities to composite orders. 
In particular, we find that the $2 N e$ pair susceptibility and $2 N k_{\mathrm{F}}$ density-wave charge/flavor susceptibilities become the most relevant at strong enough attractive and repulsive interactions, respectively.
For example, in case of $N = 2$ (spin-degenerate electron gas) we expect composite $4e$ superconductivity for large enough attractive interactions, and composite $4 k_{\mathrm{F}}$ charge/spin density waves for large enough repulsive interactions. 
The presented microscopic theory of the formation of such composite orders may provide valuable theoretical insight into the vestigial order theories commonly used to describe intertwined phase diagrams of HTSCs \cite{agterbergPhysicsPairDensityWaves2020,bergCharge4eSuperconductivityPairdensitywave2009,bergStripedSuperconductorsHow2009,agterbergCheckerboardOrderVortex2015,wangPairDensityWaves2018,fernandes_charge-_2021,samoilenka_microscopic_2026}.

Our work is organized as follows. In Sec.~\ref{sec:twopoint}, we provide the dimensional reduction map for charge, flavor and pair susceptibilities as well as for the composite susceptibilities.
Effective 1D interaction is expressed in terms of the physical $D$-dimensional RPA interaction in Sec.~\ref{sec:1Dint}.
Detailed derivation of 1D two-point correlation functions for arbitrary forward-scattering interactions is presented in Sec.~\ref{sec:boso}.
Dangers of literal application of the multidimensional bosonization are demonstrated on single-particle Green function in Sec.~\ref{sec:GF}.
The validity range and necessary regularization of the multidimensional bosonization procedure are discussed in Sec.~\ref{sec:bosoreg}.
Charge, flavor, and pair susceptibilities are investigated in Sec.~\ref{sec:susc}.
Composite order susceptibilities are studied in Sec.~\ref{sec:higherorder}.
Conclusions are given in Sec.~\ref{sec:conclusions}.
Technical details on the dimensional reduction, RPA interaction, and an example of $D$-dimensional Luttinger liquids are outlined in Appendices~\ref{sec:dimred}, \ref{sec:RPA}, and \ref{sec:LL}, respectively.

\section{Two-point correlation functions}
\label{sec:twopoint}

In this section we apply the dimensional reduction to express $D$-dimensional two-point correlation functions in terms of the effective 1D correlators.
If the $D$-dimensional density-density interaction $U_{\mathrm{D}}(\tau, r)$ [$\tau$ is the imaginary time, $r = |\bm r|$, $\bm r$ is the $D$-dimensional coordinate] transfers small momentum $p = |\bm p| \ll k_{\mathrm{F}}$ and small frequency $\omega_n \ll E_{\mathrm{F}}$ (this is what we call the forward-scattering interaction here), the dimensional reduction procedure maps $D$-dimensional fermion lines onto the corresponding 1D fermion lines via Eq.~(\ref{Fdimred}), see derivation in Appendix~\ref{sec:dimred} that is based on results of Ref.~\cite{miserevDimensionalReductionLuttingerWard2023a}.
For symmetrized closed fermion lines (fermion loops) with small-momentum and small-frequency interaction vertices, the 1D fermion loop cancellation theorem applies: the only non-zero symmetrized 1D fermion loop is the particle-hole bubble \cite{neumayr_fermion_1998}. 
In particular, this justifies the RPA approximation for dressed forward-scattering interaction $U_{\mathrm{D}}(\tau, r)$ which is equivalent to saying that the polarization operator $\Pi_{\mathrm{D}}(\tau, r)$ is represented by the particle-hole bubble \cite{frohlichEffectiveGaugeField1995,metznerFermiSystemsStrong1998,KwonHoughtonMarston1995TheoryFermionLiquids}, see Appendix~\ref{sec:RPA}.
Therefore, the dimensional reduction allows the following 1D maps: (i) diagrams with the RPA forward-scattering interaction connecting open fermion lines are mapped onto corresponding 1D diagrams via Eq.~(\ref{Fdimred}); (ii) diagrams containing fermion loops with more than two forward-scattering interaction vertices are omitted due to the loop cancellation theorem \cite{neumayr_fermion_1998}; (iii) dressing by the forward-scattering polarization operator (particle-hole bubble) must be accounted for before the dimensional reduction is applied.
As an example, we consider the fermion Green function, spin, charge, and pair susceptibilities, as well as composite correlators such as the $4 e$ pair susceptibility and $4 k_{\mathrm{F}}$ density-wave susceptibility.

The single-particle fermion Green function is connected to the corresponding 1D fermion Green function by Eq.~(\ref{Gnu}) in Appendix~\ref{sec:dimred}. 
This asymptotic form has been used before in Refs.~\cite{miserevDimensionalReductionLuttingerWard2023a,laksonoSingletSuperconductivityEnhanced2018,miserev_instability_2022}.
In the case of a non-spherical Fermi surface, the asymptotic form of the Green function is derived in Ref.~\cite{miserev_instability_2022}.
In this paper, we only consider a spherical Fermi surface.

The pair susceptibility $\chi_{\mathrm{P}}(\tau, r)$ is given by the sum of the diagrams containing two fermion lines with the forward-scattering $D$-dimensional RPA interaction $U_{\mathrm{D}}(\tau, r)$.
Applying Eq.~(\ref{Fdimred}) to both fermion lines results in the following expression for the pair susceptibility,
\begin{eqnarray}
	&& \chi_{\mathrm{P}}(\tau, r) = \sum\limits_{\nu, \nu'} \frac{e^{i(\nu + \nu')(k_{\mathrm{F}} r - \vartheta_{\mathrm{D}})}}{\left(\lambda_{\mathrm{F}} r\right)^{D - 1}} \chi^{(1D)}_{\mathrm{P}, \nu\nu'}(\tau, r) \, , \label{chiP1}
\end{eqnarray}
where $\chi_{\mathrm{P}, \nu\nu'}^{(1D)}(\tau, r)$ is the 1D pair susceptibility with chirality indices $\nu$ and $\nu'$ along the fermion lines, $\lambda_{\mathrm{F}} = 2 \pi/k_{\mathrm{F}}$ the Fermi wavelength, $\vartheta_{\mathrm{D}}$ the semiclassical phase given by Eq.~(\ref{thetaphase}).
We point out that the $2 k_{\mathrm{F}}$ contribution does not develop non-analyticities as all 1D Green functions have the same chirality index $\nu' = \nu$, so all of them are analytic on the same side of the complex frequency plane. 
The remaining contribution corresponds to $\nu' = -\nu$ yielding familiar $p \sim 0$ pair susceptibility,
\begin{eqnarray}
	&& \chi_{\mathrm{P}}(\tau, r)\left.\right|_{p \sim 0} = \frac{1}{\left(\lambda_{\mathrm{F}} r\right)^{D - 1}} \sum\limits_{\nu}  \chi^{(1D)}_{\mathrm{P}, -\nu\nu}(\tau, r) \, . \label{chiPq0}
\end{eqnarray}
We also call $\chi_{\mathrm{P}}(\tau, r)\left.\right|_{p \sim 0}$ the Cooper susceptibility.

Very similar expressions are valid for $2 k_{\mathrm{F}}$ spin/flavor and charge susceptibilities as all the diagrams consist of two fermion lines (one is for particles, another one -- for holes) connected by forward-scattering RPA interaction lines (here, we set $2 k_{\mathrm{F}}$ interaction matrix element to zero, as it does not represent the forward scattering),
\begin{eqnarray}
	&& \hspace{-25pt} \chi_{\mathrm{C/S}}(\tau, r)\left.\right|_{2k_{\mathrm{F}}} = \sum\limits_{\nu} \frac{e^{-2i \nu(k_{\mathrm{F}} r - \vartheta_{\mathrm{D}})}}{\left(\lambda_{\mathrm{F}} r\right)^{D - 1}} \chi^{(1D)}_{\mathrm{C/S}, -\nu\nu}(\tau, r) \, , \label{chiC2kF}
\end{eqnarray}
where $\chi^{(1D)}_{\mathrm{C/S}, -\nu\nu}(\tau, r)$ is the 1D charge/spin susceptibility where ``particles'' (``holes'') have chirality $-\nu$ ($\nu$).
We stress that $2k_{\mathrm{F}}$ charge and spin/flavor susceptibilities are given by the same expression as $2 k_{\mathrm{F}}$ matrix element of the forward-scattering interaction is zero [the $2 k_{\mathrm{F}}$ annihilation channel for the charge susceptibility is excluded].
At $p \sim 0$, the spin/flavor susceptibility is still represented by diagrams with two fermion lines connected by the RPA forward-scattering interaction, hence the dimensional reduction map [see Eq.~(\ref{Fdimred})] can be applied directly,
\begin{eqnarray}
	&& \hspace{-25pt} \chi_{\mathrm{S}}(\tau, r)\left.\right|_{p \sim 0} = \sum\limits_{\nu} \frac{1}{\left(\lambda_{\mathrm{F}} r\right)^{D - 1}} \chi^{(1D)}_{\mathrm{S}, \nu\nu}(\tau, r) \, . \label{chiS0}
\end{eqnarray}
In case of $p \sim 0$ charge susceptibility, we also have to sum over the annihilation channel representing the RPA series for the density-density correlator. 
As all the vertices correspond to $p \sim 0$ charge operator, the loop cancellation theorem applies directly, and $p \sim 0$ charge susceptibility is given by the RPA expression,
\begin{eqnarray}
	&& \hspace{-20pt} \chi_{\mathrm{C}}(i \omega_n, p)\left.\right|_{p \sim 0} = \frac{-\Pi_{\mathrm{D}}(i \omega_n, p)}{1 - \Pi_{\mathrm{D}}(i \omega_n, p) U_0(i \omega_n, p)} \, , \label{chiC0}
\end{eqnarray}
where $p = |\bm p| \ll k_{\mathrm{F}}$, $|\omega_n| \ll E_{\mathrm{F}}$, $\Pi_{\mathrm{D}}(i \omega_n, p)$ is the $D$-dimensional particle-hole bubble [see Eq.~(\ref{PiDhyper})], $U_0(i \omega_n, p)$ is the bare $D$-dimensional interaction connected with the RPA interaction $U_{\mathrm{D}}(i \omega_n, p)$ via standard Dyson equation Eq.~(\ref{URPA}), see Appendix~\ref{sec:RPA} for details.
The sign convention for $\chi_{\mathrm{C}}(i \omega_n, p)\left.\right|_{p \sim 0}$ corresponds to positive static density-density correlator in case if the Fermi-liquid ground state is stable.
Similarly, the $p \sim 0$ spin/flavor susceptibility [see Eq.~(\ref{chiS0})] can be directly expressed in terms of the $D$-dimensional particle-hole bubble,
\begin{eqnarray}
	&& \chi_{\mathrm{S}}(i \omega_n, p)\left.\right|_{p \sim 0} \approx -\Pi_{\mathrm{D}}(i \omega_n, p) \, , \label{chiSPi}
\end{eqnarray}
where the spin/flavor dependence is ``traced out'' in $SU(N)$ symmetric system, $N$ is the number of fermion flavors.
We stress that Eqs.~(\ref{chiC0}), (\ref{chiSPi}) take into account the leading non-analyticity defining the long-distance asymptotics of these functions.
In particular, analytic part of the polarization operator that does not contribute to the long-distance asymptotics of $\Pi_{\mathrm{D}}(\tau, r)$ are not taken into account.
For example, these analytic corrections to $\Pi_{\mathrm{D}}(\tau, r)$ arise from fermion loops in the full $D$-dimensional theory that we neglected in our long-distance approximation.
One important consequence of this is the RPA screening length $R_{\mathrm{s}}$ has to be considered as a phenomenological parameter of the theory.

In case of $N$ fermion species, we may construct composite susceptibilities containing $2N$ fermion lines,
\begin{eqnarray}
	&& \hspace{-20pt} \chi_{\mathrm{P}, \mathrm{2N}}(\tau, r) = \sum\limits_{\{\nu_j\}} \frac{e^{i(k_{\mathrm{F}} r - \vartheta_{\mathrm{D}}) \sum\limits_{j = 1}^{2 N} \nu_j}}{\left(\lambda_{\mathrm{F}} r\right)^{N(D - 1)}} \chi^{(1D)}_{\mathrm{P}, \mathrm{2N}, \{\nu_j\}}(\tau, r) \, , \label{chiPN} \\
	&& \hspace{-20pt} \chi_{\mathrm{C/S}, \mathrm{2N}}(\tau, r) \nonumber \\
	&& \hspace{-20pt} = \sum\limits_{\{\nu_j,\nu'_j\}} \frac{e^{i\sum\limits_{j = 1}^N (\nu_j - \nu'_j)(k_{\mathrm{F}} r - \vartheta_{\mathrm{D}})}}{\left(\lambda_{\mathrm{F}} r\right)^{N(D - 1)}} \chi^{(1D)}_{\mathrm{C/S}, \mathrm{2N}, \{\nu_j, \nu'_j\}}(\tau, r) \, , \label{chiCSN}
\end{eqnarray}
where the susceptibilities with the subscript $^{(1D)}$ correspond to the 1D susceptibilities with the same convolution rules of the flavor/spin matrices.
For charge and spin susceptibilities indices $\{\nu_j\}$ ($\{\nu_j'\}$) correspond to ``particles'' (``holes'').
We specifically consider vertex operators for flavor/charge susceptibilities that do not contain $p \sim 0$ flavor density operators of the form $\overline{\Psi}_{\alpha}(\bm k) \Psi_\alpha (\bm k + \bm p)$ in order to avoid the RPA-like summation, and apply Eq.~(\ref{chiCSN}) directly.
This also makes sense as the contribution of $p \sim 0$ flavor density operators is non-singular.
We discuss the most relevant harmonics of each composite susceptibility in Sec.~\ref{sec:higherorder}.
If $N = 2$ [which corresponds to the spin-degenerate electron gas], the pair susceptibility $\chi_{\mathrm{P}, \mathrm{4}}(\tau, r)$ probes the $4e$ superconducting instability, where the charged collective mode is made of four electrons.
Hence, our theory provides a clear microscopic mechanism of composite 4e superconductivity which is usually treated phenomenologically within the effective Ginzburg-Landau theory of vestigial orders \cite{agterbergPhysicsPairDensityWaves2020,bergCharge4eSuperconductivityPairdensitywave2009,bergStripedSuperconductorsHow2009,agterbergCheckerboardOrderVortex2015,wangPairDensityWaves2018,fernandes_charge-_2021,samoilenka_microscopic_2026}.
This phenomenon can be measured in quantum interference experiments capturing the charge of such composite states \cite{huang_evidence_2022,ge_charge-4e_2024,almoalem_observation_2024}.
However, quantum oscillation experiments alone are generally not sufficient for an undeniable proof of composite superconductivity due to other possible explanations such as the Higgs-Leggett mechanism in multi-component systems \cite{varma_extended_2023,zhang_higgs-leggett_2024}.
The $4e$ superconductivity is also predicted numerically in $t-J$ and Hubbard models in the strong coupling regime \cite{soldini_charge-_2024,golic_demonstration_2026,shi_high-temperature_2026}.

\section{Effective 1D Interaction}
\label{sec:1Dint}

In this section we construct the effective 1D interaction that we use in order to calculate the 1D Green function and 1D susceptibilities introduced in Sec.~\ref{sec:twopoint}.

In the dimensional reduction, the reduced diagrams contain dressed interaction $U_{\mathrm{D}}(\tau, r)$ that is approximated via RPA, see Eq.~(\ref{URPA}) in Appendix~\ref{sec:RPA}.
As we mapped two-point correlation functions onto their 1D analogues, it is natural to apply the bosonization technique to calculate them.
In order to do so, we have to restore the bare effective 1D interaction $V_0(\tau, x)$.
We already found that the dressed 1D interaction $V(\tau, x)$ simply equals $U_{\mathrm{D}}(\tau, |x|)$, 
\begin{eqnarray}
	&& V(\tau, x) = U_{\mathrm{D}}(\tau, |x|) \, , \label{UV} \\
	&& V(i \omega_n, q) = \int\limits_{-\infty}^\infty U_{\mathrm{D}}(i \omega_n, |x|) e^{-i q x} \, dx \, , \label{V1Da} \\
	&& U_{\mathrm{D}}(i \omega_n, r) = \int \frac{d \bm p}{(2 \pi)^D}  e^{i \bm p \cdot \bm r} U_{\mathrm{D}}(i \omega_n, p)  \, , \label{UDa}
\end{eqnarray}
where $V(i \omega_n, q)$ is the 1D Fourier transform of $V(\tau, x)$.
As the RPA approximation is asymptotically exact for any $D \ge 1$, the bare 1D interaction $V_0(i \omega_n, q)$ follows from the standard Dyson equation where the polarization operator is given by the bare particle-hole bubble $\Pi_{\mathrm{1D}}^{(0)}(i \omega_n, q)$,
\begin{eqnarray}
	&& \hspace{-15pt} V_0(i \omega_n, q) = \left(V^{-1}(i \omega_n, q) +  \Pi_{\mathrm{1D}}^{(0)}(i \omega_n, q) \right)^{-1} \, , \label{Vgen} \\
	&& \hspace{-15pt} \Pi_{\mathrm{1D}}^{(0)}(i \omega_n, q) = - \frac{N}{\pi v_{\mathrm{F}}} \frac{(q v_{\mathrm{F}})^2}{\omega_n^2 + (q v_{\mathrm{F}})^2} \, . \label{dPi1D}
\end{eqnarray}

Next, we derive a convenient representation of the dressed 1D interaction in terms of the $D$-dimensional RPA interaction.
For this, we substitute Eq.~(\ref{UDa}) into Eq.~(\ref{V1Da}) and use the expression for the angular integral given by Eq.~(\ref{angleint}), 
\begin{eqnarray}
	&& \hspace{-20pt} V(i \omega_n, q) = \int\limits_0^\infty U_{\mathrm{D}}(i \omega_n, p) I_{\mathrm{D}}\left(\frac{q}{p}\right) \frac{p^{D - 2} \, dp}{(2 \pi)^{\frac{D}{2}}} \, , \label{VID}\\
	&& \hspace{-20pt} I_{\mathrm{D}}(z) = \int\limits_{-\infty}^\infty J_{\frac{D}{2} - 1} (|y|) e^{- i z y} \frac{dy}{|y|^{\frac{D}{2} - 1}} \, .
\end{eqnarray}
In order to evaluate $I_{\mathrm{D}}(z)$, we use the integral representation of Bessel functions given by Eq.~(\ref{Bessel}),
\begin{eqnarray}
	&& I_{\mathrm{D}}(z) = \frac{\sqrt{\pi} \theta_{\mathrm{H}}(1 - |z|)}{2^{\frac{D}{2} - 2} \Gamma\left(\frac{D-1}{2}\right)} \left(1 - z^2\right)^{\frac{D - 3}{2}} \, ,
\end{eqnarray}
where $\theta_{\mathrm{H}}(z)$ is the Heaviside step function.
This leads to the following expression for the dressed 1D interaction,
\begin{eqnarray}
	&& V(i \omega_n, q) = \frac{|q|^{D - 1}}{2^{D - 2} \pi^{\frac{D-1}{2}} \Gamma\left(\frac{D-1}{2}\right)} \nonumber \\
	&& \times \int\limits_{1}^\infty U_{\mathrm{D}} (i \omega_n, |q| z) \left(z^2 -1\right)^{\frac{D-3}{2}} z \, dz \, . \label{V1D}
\end{eqnarray}
We point out that in the limit $D \to 1$ Eq.~(\ref{V1D}) simply gives $V(i \omega_n, q) = U_{\mathrm{D}} (i \omega_n, |q|)$.
Equation~(\ref{V1D}) agrees with the one derived in Ref.~\cite{metznerFermiSystemsStrong1998}.

\section{1D Bosonization with Arbitrary Forward Scattering}
\label{sec:boso}

In this section we perform the 1D bosonization with arbitrary 1D interaction $V_0(i \omega_n, q)$.
We then apply it to calculate the 1D fermion Green function and 1D susceptibilities introduced in Sec.~\ref{sec:twopoint}.

In this section, $\omega_n = 2 \pi T n$ is the bosonic Matsubara frequency, $T$ the temperature, $n$ an integer, $q \ll k_{\mathrm{F}}$ the momentum transfer. We assume that all $N$ fermion species have the same Fermi velocity $v_{\mathrm{F}}$.
In the low-energy limit, the 1D system is described by the following action,
\begin{eqnarray}
	&& \hspace{-25pt} \mathcal{S} = \mathcal{S}_0 + \mathcal{S}_{\mathrm{int}} \, , \label{action}\\
	&& \hspace{-25pt} \mathcal{S}_0 = \sum\limits_{\alpha = 1}^N \int d\xi \, \left[-i \Pi_{\alpha}(\xi) \partial_\tau \varphi_{\alpha}(\xi) \vphantom{\frac{v_{\mathrm{F}}}{2}} \right. \nonumber \\
	&& \left. + \frac{v_{\mathrm{F}}}{2} \left(\Pi_{\alpha}^2(\xi) + \left(\partial_x \varphi_{\alpha}(\xi)\right)^2\right) \right] \, , \\
	&& \hspace{-25pt} \mathcal{S}_{\mathrm{int}} = \sum\limits_{\alpha, \alpha'} \int \frac{d \xi d \xi'}{2 \pi} \, V_0(\xi - \xi') \partial_x \varphi_{\alpha}(\xi) \partial_{x'} \varphi_{\alpha'}(\xi') \, ,
\end{eqnarray}
where $\alpha, \alpha' \in \{1, \dots, N\}$ are the fermion flavor indices, $\xi = (\tau, x)$, $\tau \in (0, \beta)$ the imaginary time, $x$ the 1D coordinate, $\beta = 1/T$, $\partial_\tau$ and $\partial_x$ are derivatives with respect to $\tau$ and $x$, $\varphi_{\alpha}(\xi)$ the bosonic field, $\Pi_{\alpha}(\xi)$ its conjugate:
\begin{eqnarray}
	&& \hspace{-15pt} \varphi_{\alpha}(x) = \sum\limits_{q \ne 0} \frac{e^{i q x}}{\sqrt{2 L |q|}} \left(a_{\alpha}(q) + a_{\alpha}^\dagger (-q)\right) \, , \\
	&& \hspace{-15pt} \Pi_{\alpha}(x) = - \partial_x \vartheta_{\alpha} (x) \nonumber \\
	&& \hspace{15pt} = -i \sum\limits_{q \ne 0} \sqrt{\frac{|q|}{2 L}} e^{i q x} \left(a_{\alpha}(q) - a_{\alpha}^\dagger (-q)\right) \, ,
\end{eqnarray}
where $L \to \infty$ is the 1D system size, $a_{\alpha}(q)$ ($a_{\alpha}^\dagger (q)$) the annihilation (creation) bosonic operators, $\vartheta_{\alpha} (x)$ is an auxiliary field.
Here, we do not consider $q = 0$ modes as their contribution vanishes in the thermodynamic limit $L \to \infty$.
The 1D fermion fields can be expanded via left and right movers as follows,
\begin{eqnarray}
	&& \Psi_{\alpha} (x) = \sum\limits_{\nu = \pm 1} e^{i \nu k_{\mathrm{F}} x} \psi_{\nu \alpha}(x) \, , \\
	&& \psi_{\nu \alpha}(x) = \mathcal{U}_{\nu \alpha} e^{i \sqrt{\pi} \left(\vartheta_{\alpha}(x) + \nu \varphi_{\alpha}(x)\right)} \, , \label{psi1D}
\end{eqnarray}
where $\mathcal{U}_{\nu \alpha}$ are Klein factors, $\nu = \pm 1$ is the chiral index, $\nu = +1$ ($\nu = -1$) corresponds to the right (left) movers. Notice that we also include the normalization in definition of the Klein factors.

The action in Eq.~(\ref{action}) can be equivalently represented as a quadratic form with respect to the bosonic fields, 
\begin{eqnarray}
	&& \hspace{-28pt} \mathcal{S}[\Phi] = \frac{T}{2L} \sum\limits_{q, \omega_n} \Phi^T(-i \omega_n, -q) M^{-1} (i \omega_n, q) \Phi(i \omega_n, q) \, , \label{SPhi}
\end{eqnarray}
where $\Phi^T(i \omega_n, q) = (\varphi(i \omega_n, q) ;  \vartheta (i \omega_n, q))$, $\varphi^T = (\varphi_{1}, \dots, \varphi_{N})$, $\vartheta^T = (\vartheta_{1}, \dots, \vartheta_{N})$, and $M^{-1}$ is the following matrix,
\begin{equation}
\frac{M^{-1}(i \omega_n, q)}{v_{\mathrm{F}} q^2} = \left(
	\begin{array}{cc}
	 \hat{I} + \frac{\displaystyle V_0(i \omega_n, q)}{\displaystyle \pi v_{\mathrm{F}}} \hat{C}  & - \frac{\displaystyle i \omega_n}{\displaystyle q v_{\mathrm{F}}} \hat{I} \\ [10pt]
	 - \frac{\displaystyle i \omega_n}{\displaystyle q v_{\mathrm{F}}} \hat{I} & \hat{I}
	\end{array}
	\right) \, ,
\end{equation}
where $\hat{I}$ is $N \times N$ identity matrix, $\hat{C}$ is the $N \times N$ matrix all entries of which are equal to one, $\hat{C}_{\alpha\alpha'} = 1$.
The matrix $M(i \omega_n, q)$ can then be expressed as follows,
\begin{equation}
	M = v_{\mathrm{F}} D(i \omega_n, q) \left(
	\begin{array}{cc}
	\hat{I} & \frac{\displaystyle i \omega_n}{\displaystyle q v_{\mathrm{F}}} \hat{I} \\ [10pt]
	 \frac{\displaystyle i \omega_n}{\displaystyle q v_{\mathrm{F}}} \hat{I} & \hat{I} + \frac{\displaystyle V_0(i \omega_n, q)}{\displaystyle \pi v_{\mathrm{F}}} \hat{C}
	\end{array}
	\right) \, , \label{M}
\end{equation}
where $D(i\omega_n, q)$ stands for the propagator of the $\varphi$ field,
\begin{eqnarray}
	&& \hspace{-30pt} D(i\omega_n, q) = \left[\left(q v_{\mathrm{F}}\right)^2 \left(\hat{I} + \frac{V_0(i \omega_n, q) \hat{C}}{\pi v_{\mathrm{F}}}\right) + \omega_n^2 \right]^{-1}  . \label{D}
\end{eqnarray}

Correlation functions can be calculated with the help of the effective action $\mathcal{S}_{\mathrm{eff}}[J]$ depending on classical ``current'' fields $J$,
\begin{eqnarray}
	&& \hspace{-23pt} \mathcal{S}_{\mathrm{eff}}[J] = - \ln \left( \int \frac{d \Phi}{Z} e^{- \mathcal{S}[\Phi] - \frac{1}{2}\left(\Phi^T J + J^T \Phi\right)} \right) \nonumber \\
	&& \hspace{-23pt} = - \frac{1}{2} J^T M J = - \frac{1}{2} \int d\xi d\xi' \, J^T(\xi) M(\xi - \xi') J(\xi') \, , \label{effaction}
\end{eqnarray}
where $Z = \int d\Phi \, e^{-\mathcal{S}[\Phi]}$ is the statistical weight.

\subsection{Spin-charge separation}

Let us expand $\varphi$ column via the charge mode $\varphi_{\mathrm{c}}$ and flavor/spin modes $\varphi_{\mathrm{s}}$, $s \in \{1, \dots, N -1\}$,
\begin{eqnarray}
	&& \hat{C} \varphi_{\mathrm{c}} = N \varphi_{\mathrm{c}} \, , \label{phiC}\\
	&& \hat{C} \varphi_{\mathrm{s}} = 0 \, . \label{phin}
\end{eqnarray}
The charge mode can be represented explicitly,
\begin{eqnarray}
	&& \varphi_{\mathrm{c}} (\xi) = \frac{1}{\sqrt{N}} \sum\limits_{\alpha = 1}^N \varphi_{\alpha} (\xi) \, . \label{charge}
\end{eqnarray}
There is a freedom to choose orthonormal flavor modes $\varphi_{\mathrm{s}}$ as $\mathrm{rank}(\hat{C}) = 1$, i.e. $\hat{C}$ has $(N - 1)$-fold degenerate eigenvalue $0$.
Explicit form of these $\varphi_{\mathrm{s}}$ modes is not important, as soon as they form an orthonormal basis.
The same orthogonal transformation can be applied to the $\vartheta$ column. The overall transformation is canonical.
We point out that Eq.~(\ref{phin}) immediately shows that the flavor/spin modes do not couple to the density-density interaction $V_0(i \omega_n, q)$.
This decoupling is a manifestation of the spin-charge separation.

Spin-charge separation results in the decoupling of the 1D action into the sum over the charge and flavor/spin sectors,
\begin{eqnarray}
	&& \hspace{-20pt} \mathcal{S}[\Phi] = \mathcal{S}_{\mathrm{c}}\left[\Phi_{\mathrm{c}}\right] + \sum\limits_{s = 1}^{N - 1} \mathcal{S}_{\mathrm{s}}\left[\Phi_{\mathrm{s}}\right] \, , \label{Sdecoupled} \\
	&& \hspace{-20pt} \mathcal{S}_{\mathrm{c}}\left[\Phi_{\mathrm{c}}\right] = \frac{1}{2} \Phi_{\mathrm{c}}^T M_{\mathrm{c}}^{-1} \Phi_{\mathrm{c}} \, , \hspace{5pt} \mathcal{S}_{\mathrm{s}}\left[\Phi_{\mathrm{s}}\right] = \frac{1}{2} \Phi_{\mathrm{s}}^T M_{\mathrm{s}}^{-1} \Phi_{\mathrm{s}} \, ,
\end{eqnarray}
where $\Phi_{\mathrm{c}}^T = (\varphi_{\mathrm{c}}, \vartheta_{\mathrm{c}})$, $\Phi_{\mathrm{s}}^T = (\varphi_{\mathrm{s}}, \vartheta_{\mathrm{s}})$, and $\Phi^T M^{-1} \Phi$ is the abbreviated form of the trace over all frequencies and momenta like in Eq.~(\ref{SPhi}).
The propagator $M_{\mathrm{s}}$ of the spin modes is completely independent of the interaction $V_0(i \omega_n, q)$,
\begin{equation}
	M_{\mathrm{s}} \equiv M_0 = v_{\mathrm{F}} D_0(i \omega_n, q) \left(
	\begin{array}{cc}
		1 & \frac{\displaystyle i \omega_n}{\displaystyle q v_{\mathrm{F}}}\\ [10pt]
		\frac{\displaystyle i \omega_n}{\displaystyle q v_{\mathrm{F}}} & 1
	\end{array}
	\right) \, , \label{Ms}
\end{equation}
where $D_0(i \omega_n, q)$ is the free-boson propagator,
\begin{eqnarray}
	&& D_0(i \omega_n, q) = \frac{1}{\omega_n^2 + (q v_{\mathrm{F}})^2} \, . \label{D0}
\end{eqnarray}
The charge modes have the renormalized propagator,
\begin{equation}
	M_{\mathrm{c}} = v_{\mathrm{F}} D_{\mathrm{c}}(i \omega_n, q) \left(
	\begin{array}{cc}
		1 & \frac{\displaystyle i \omega_n}{\displaystyle q v_{\mathrm{F}}} \\ [10pt]
		\frac{\displaystyle i \omega_n}{\displaystyle q v_{\mathrm{F}}} & 1 + \frac{\displaystyle N V_0(i \omega_n, q)}{\displaystyle \pi v_{\mathrm{F}}}
	\end{array}
	\right) \, , \label{Mc}
\end{equation}
where $ D_{\mathrm{c}}(i \omega_n, q)$ is the propagator of the charge mode $\varphi_{\mathrm{c}}$,
\begin{eqnarray}
	&& \hspace{-30pt} D_{\mathrm{c}}(i\omega_n, q) = \left[\left(q v_{\mathrm{F}}\right)^2 \left(1 + \frac{N V_0(i \omega_n, q)}{\pi v_{\mathrm{F}}}\right) + \omega_n^2 \right]^{-1} \,  . \label{Dc}
\end{eqnarray}
Using Eqs.~(\ref{Vgen}), (\ref{dPi1D}), we can represent $D_{\mathrm{c}}(i\omega_n, q)$ in terms of dressed 1D interaction $V(i \omega_n, q)$ as follows,
\begin{eqnarray}
	&& D_{\mathrm{c}}(i\omega_n, q) = D_0 (i\omega_n, q) \frac{V(i \omega_n, q)}{V_0(i \omega_n, q)} \, . \label{DcVV0}
\end{eqnarray}
The effective action $\mathcal{S}_{\mathrm{eff}}[J]$ is then represented as the sum over the charge and spin sectors as follows,
\begin{eqnarray}
	&& \mathcal{S}_{\mathrm{eff}}[J] = -\frac{1}{2} J_{\mathrm{c}}^T M_{\mathrm{c}} J_{\mathrm{c}} -\frac{1}{2} \sum\limits_{s = 1}^{N - 1} J_{\mathrm{s}}^T M_0 J_{\mathrm{s}} \, . \label{Seffseparated}
\end{eqnarray}
Here, $J_{\mathrm{c}}$ and $J_{\mathrm{s}}$ are charge and spin components of the current $J = (J^{(\varphi)}, J^{(\vartheta)})^T$ with $J^{(\varphi)}$ ($J^{(\vartheta)}$) being the $\varphi$-component ($\vartheta$-component) of the vector $J$,
\begin{equation}
	J_{\mathrm{c}} = \left(
	\begin{array}{c}
		u_{\mathrm{c}}^T J^{(\varphi)} \\
		u_{\mathrm{c}}^T J^{(\vartheta)}
	\end{array} \right) \, , \hspace{5pt} 
	J_{\mathrm{s}} = \left(
	\begin{array}{c}
		u_{\mathrm{s}}^T J^{(\varphi)} \\
		u_{\mathrm{s}}^T J^{(\vartheta)}
	\end{array} \right) \, , \label{JcJs}
\end{equation}
where $u_{\mathrm{c}}$ and $u_{\mathrm{s}}$ are orthonormal eigenvectors of the $N \times N$ matrix $\hat{C}$,
\begin{eqnarray}
	&& \hat{C} u_{\mathrm{c}} = N u_{\mathrm{c}} \, , \hspace{5pt} \hat{C} u_{\mathrm{s}} = 0 \, .
\end{eqnarray}
As the spin modes remain unrenormalized by the interaction $V_0(i \omega_n, q)$, they are not of interest for us: those effects can be included into the non-interacting part of the correlation function of our interest.
The charge current $J_c$, on the other hand, is important, so we provide explicit expression for it,
\begin{equation}
	J_{\mathrm{c}} \equiv \left(
	\begin{array}{c}
		J_{\mathrm{c}}^{(\varphi)} \\
		J_{\mathrm{c}}^{(\vartheta)}
	\end{array} \right) = \frac{1}{\sqrt{N}}\left(
	\begin{array}{c}
		\mathrm{tr} \left[J^{(\varphi)}\right] \\
		\mathrm{tr} \left[J^{(\vartheta)}\right]
	\end{array} \right) \, , \label{Jcexplicit} 
\end{equation}
where $\mathrm{tr}[J^{\varphi}]$ stands for the sum of all $N$ entries of $J^{\varphi}$, and similarly for $\mathrm{tr}[J^{\vartheta}]$.

\subsection{Two-point correlation functions}

We aim to calculate two-point correlation functions $\chi_{\mathrm{A\overline{A}}}(\xi - \xi')$ with vertex operators $A(\xi)$, $\overline{A}(\xi')$ [the conjugate of $A(\xi')$] that are polynomials in the fermion field operators,
\begin{eqnarray}
	&& A(\xi) \propto \overline{\psi}_{\nu_1 \alpha_1} \dots \overline{\psi}_{\nu_n \alpha_n} \psi_{\nu_1' \alpha_1'} \dots \psi_{\nu_m' \alpha_m'} (\xi) \, , \label{A}
\end{eqnarray}
where $n, m \in \{1, \dots, N\}$, all fermion operators are taken at $\xi = (\tau, x)$, $\overline{\psi}_{\nu \alpha}(\xi)$ is the conjugate fermion field. 
For simplicity, we consider all $\psi$ (all $\overline{\psi}$) operators different, i.e. $\{\nu_j, \alpha_j\} \ne \{\nu_i, \alpha_i\}$ ($\{\nu'_j, \alpha'_j\} \ne \{\nu'_i, \alpha'_i\}$) at $i \ne j$.
Moreover, we assume no $q \sim 0$ flavor density operators $\overline{\psi}_{\nu \alpha} (\xi) \psi_{\nu \alpha}(\xi)$, i.e. no $\{\nu_i', \alpha_i'\}$ is equal to $\{\nu_j, \alpha_j\}$.
In this case, we can directly use the exponential map for 1D fermion operators [see Eq.~(\ref{psi1D})] without worrying about point splitting, Klein factors, and 1D RPA summation.
We use the functional integral representation of $\chi_{\mathrm{A\overline{A}}}(\xi - \xi')$,
\begin{eqnarray}
	&& \chi_{\mathrm{A\overline{A}}}(\xi - \xi') = \int \frac{d \Phi}{Z} \, e^{-\mathcal{S}[\Phi]} A(\xi) \overline{A}(\xi') \, . \label{chiAA}
\end{eqnarray}
The Klein factors are evaluated the same way as for the non-interacting correlation function.
It is convenient to use the bosonized representation Eq.~(\ref{psi1D}) for the fermion fields, leading to an expression of the form of Eq.~(\ref{effaction}).
Pulling the non-interacting correlation function $\chi^{(0)}_{\mathrm{A\overline{A}}}(\xi - \xi')$ out of this expression yields the following result,
\begin{eqnarray}
	&& \hspace{-25pt} \chi_{\mathrm{A\overline{A}}}(\xi - \xi') = \chi^{(0)}_{\mathrm{A\overline{A}}}(\xi - \xi') F_{\mathrm{A\overline{A}}}(\xi - \xi') \, , \\
	&& \hspace{-25pt} F_{\mathrm{A\overline{A}}}(\xi - \xi') \nonumber \\
	&& \hspace{-25pt} = \exp\left(\frac{1}{2} \int d\xi_1 d\xi_2 \, J_{\mathrm{c}}^T(\xi_1) \delta M_{\mathrm{c}} (\xi_1 - \xi_2) J_{\mathrm{c}} (\xi_2)\right) \, ,  \label{Ffactor} \\
	&& \hspace{-25pt} \delta M_c (\xi) = M_c(\xi) - M_0(\xi) \, , \label{dM}
\end{eqnarray}
where $M_0(\xi)$ and $M_c(\xi)$ are 1D Fourier transforms of bosonic propagators given by Eqs.~(\ref{Ms}), (\ref{Mc}), respectively.
The current $J$ comes from the fermion fields, see Eq.~(\ref{psi1D}), so it is inserted in Eq.~(\ref{chiAA}) at points $\xi$ and $\xi'$,
\begin{eqnarray}
	&& J(\xi_1) = J_{\mathrm{A}} \delta(\xi_1 - \xi) + J_{\mathrm{\overline{A}}} \delta(\xi_1 - \xi') \, ,
\end{eqnarray}
where $J_{\mathrm{\overline{A}}} = - J_{\mathrm{A}}$. 
The charge component of $J_{\mathrm{A}}$ is the following,
\begin{equation}
	J_{\mathrm{A}, \mathrm{c}} \equiv i \sqrt{\frac{\pi}{N}} \, j_{\mathrm{A}, \mathrm{c}} \, , \hspace{5pt} j_{\mathrm{A}, \mathrm{c}} = \left(
	\begin{array}{c}
		\mathcal{J}_{\mathrm{A}} \\[3pt]
		\mathcal{Q}_{\mathrm{A}}
	\end{array}
	\right) \, , \label{JAc}
\end{equation}
where integers $\mathcal{Q}_{\mathrm{A}}$ and $\mathcal{J}_{\mathrm{A}}$ correspond to the charge and the current quantum numbers associated with the operator $A(\xi)$, see Eq.~(\ref{A}),
\begin{eqnarray}
	&& \mathcal{Q}_{\mathrm{A}} = m - n \, , \hspace{5pt} \mathcal{J}_{\mathrm{A}} = \sum\limits_{j = 1}^m \nu_j' - \sum\limits_{j = 1}^n \nu_j  \, . \label{QJ}
\end{eqnarray}
The susceptibility is then simplified to the following expression,
\begin{eqnarray}
	&& \hspace{-30pt} \frac{\chi_{\mathrm{A\overline{A}}}(\xi)}{\chi^{(0)}_{\mathrm{A\overline{A}}}(\xi)} =  \exp\left(-j_{\mathrm{A}, \mathrm{c}}^T \left(M_{\mathrm{int}}\left(\xi\right) - M_{\mathrm{int}}\left(0\right) \right) j_{\mathrm{A}, \mathrm{c}}\right) \, , \label{chiAAfinal}
\end{eqnarray} 
where we used that $J_{\mathrm{\overline{A}}} = - J_{\mathrm{A}}$, $\delta M_{\mathrm{c}}(-\xi) = \delta M_{\mathrm{c}}(\xi)$, $j_{\mathrm{A}, \mathrm{c}}$ is given by Eq.~(\ref{JAc}), and $M_{\mathrm{int}}(i \omega_n, q)$ is the following matrix,
\begin{eqnarray}
	&& \hspace{-30pt} M_{\mathrm{int}} (i \omega_n, q) \equiv - \frac{\pi}{N} \delta M_{\mathrm{c}} (i \omega_n, q) \nonumber \\
	&& \hspace{-30pt} = V(i \omega_n, q) D_0^2(i \omega_n, q) \left(
	\begin{array}{cc}
		\left(q v_{\mathrm{F}}\right)^2 & i \omega_n q v_{\mathrm{F}} \\
		i \omega_n q v_{\mathrm{F}} & - \omega_n^2
	\end{array}
	\right) \, , \label{dMc}
\end{eqnarray} 
where $V(i \omega_n, q)$ is the dressed 1D interaction, expressed in terms of the $D$-dimensional RPA interaction via Eq.~(\ref{V1D}).
We point out that Eq.~(\ref{chiAAfinal}) is true for any metallic 1D fermion system with any forward-scattering interaction $V_0(i \omega_n, q)$. In particular, if $V_0(i \omega_n, q) = const$, this result corresponds to the Luttinger liquid \cite{giamarchi}.

In our case, the dressed 1D interaction is directly connected to the $D$-dimensional RPA interaction via the correspondence given by Eq.~(\ref{UV}).
Using the linearity of $M_{\mathrm{int}}(i \omega_n, q)$ with respect to the dressed 1D interaction $V(i \omega_n, q)$, we can represent $M_{\mathrm{int}}(\xi)$ as the following convolution,
\begin{eqnarray}
	&& M_{\mathrm{int}}(\xi) = \int d\xi' \, U_{\mathrm{D}}(\xi') \mathcal{K}_{\mathrm{int}}(\xi - \xi') \, , \label{Mint}
\end{eqnarray}
where $\xi = (\tau, x)$, and we used Eq.~(\ref{UV}).
The kernel $\mathcal{K}_{\mathrm{int}}(\xi)$ and corresponding two-point correlation function are given by the following expressions,
\begin{widetext}
\begin{eqnarray}
	&& \mathcal{K}_{\mathrm{int}}(\xi) = -\frac{\ln\left|2 \sin \left(\pi T z\right)\right|}{4 \pi v_{\mathrm{F}}} \left(
	\begin{array}{cc}
		1 & 0 \\
		0 & -1
	\end{array}
	\right)  - \frac{T x}{8 v_{\mathrm{F}}^2 \left|\sin \left(\pi T z\right)\right|^2} \left(
	\begin{array}{cc}
		\sinh\left(2 \pi T \frac{x}{v_{\mathrm{F}}}\right) & - i \sin\left(2 \pi T \tau\right) \\ [3pt]
		- i \sin\left(2 \pi T \tau\right) & \sinh\left(2 \pi T \frac{x}{v_{\mathrm{F}}}\right)
	\end{array}
	\right) \, , \label{Kint} \\ [5pt]
	&& \chi_{\mathrm{A\overline{A}}}(\xi) = \chi^{(0)}_{\mathrm{A\overline{A}}}(\xi) \exp\left[-\int d\xi' \, U_{\mathrm{D}}(\xi') \, j_{\mathrm{A}, \mathrm{c}}^T \left(\mathcal{K}_{\mathrm{int}}\left(\xi - \xi'\right) - \mathcal{K}_{\mathrm{int}}\left( - \xi'\right)\right) j_{\mathrm{A}, \mathrm{c}}\right] \, , \label{chiAA1D}
\end{eqnarray}
\end{widetext}
where $z = \tau - i x/v_{\mathrm{F}}$.
In order to evaluate $\mathcal{K}_{\mathrm{int}}(\xi)$, we used the following identities,
\begin{eqnarray}
	&& \hspace{-29pt} \int\limits_{-\infty}^\infty \frac{dq}{2 \pi} \, \frac{e^{i q x}}{\left(\omega_n^2 + (q v_{\mathrm{F}})^2\right)^2} = \frac{e^{-\frac{|\omega_n x|}{v_{\mathrm{F}}}}}{4 v_{\mathrm{F}} |\omega_n|^3} \left(1 + \frac{|\omega_n x|}{v_{\mathrm{F}}}\right) \, , \\
	&& \hspace{-29pt} \int\limits_{-\infty}^\infty \frac{dq}{2 \pi} \, \frac{q^2 e^{i q x}}{\left(\omega_n^2 + (q v_{\mathrm{F}})^2\right)^2} = \frac{e^{-\frac{|\omega_n x|}{v_{\mathrm{F}}}}}{4 v_{\mathrm{F}}^3 |\omega_n|} \left(1 - \frac{|\omega_n x|}{v_{\mathrm{F}}}\right) \, , \\ 
	&& \hspace{-29pt} T \sum\limits_{\omega_n \ne 0} e^{- i \omega_n \tau - \frac{|\omega_n x|}{v_{\mathrm{F}}}} = \frac{T \sinh\left(2 \pi T \frac{|x|}{v_{\mathrm{F}}}\right)}{2 \left|\sin\left(\pi T z\right)\right|^2} - T \, , \\
	&& \hspace{-29pt} T \sum\limits_{\omega_n \ne 0} \frac{e^{- i \omega_n \tau}}{|\omega_n|} e^{- \frac{|\omega_n x|}{v_{\mathrm{F}}}} = \frac{T |x|}{v_{\mathrm{F}}} - \frac{1}{\pi} \ln\left|2 \sin\left(\pi T z\right)\right| \, . \label{wsum2}
\end{eqnarray}
We point out that the Fourier transform of $q^2 D_0^2(i \omega_n, q)$ contains a divergent contribution coming from the zero Matsubara frequency $\omega_n = 0$ leading to the integral of the following form,
\begin{eqnarray}
	&& \int\limits_{-\infty}^\infty \frac{d q}{q^2} \, e^{i q x} = \int\limits_{-\infty}^\infty \frac{\cos\left(q x\right) - 1}{q^2} \, dq + \int\limits_{-\infty}^\infty \frac{dq}{q^2} \nonumber \\
	&& = - \pi \left|x\right| + \mathrm{const.} \, ,
\end{eqnarray}
where we explicitly separated the divergent constant term. 
This term is canceled out by the same constant contribution from $M_{\mathrm{int}}(0)$ in Eq.~(\ref{chiAAfinal}) leaving Eq.~(\ref{chiAA1D}) free of infinities.

\section{Fermion Green function}
\label{sec:GF}

In case of the fermion Green function, the operator $A(\xi)$ is simply given by $\psi_{\nu \alpha}(\xi)$, so $\mathcal{J}_{\mathrm{A}} = \nu$, $\mathcal{Q}_{\mathrm{A}} = 1$ yielding $j_{\mathrm{A}, \mathrm{c}}^T = (\nu, 1)$.
Substituting this into Eq.~(\ref{chiAA1D}) and using Eq.~(\ref{g0taux}) for the non-interacting Green function, we find,
\begin{eqnarray}
	&& g_{\nu \alpha} (\xi) = - \frac{\displaystyle T \exp\left[\mathcal{L}_\nu(\xi)\right]}{2 v_{\mathrm{F}} \displaystyle \sin\left(\pi T z_\nu\right)} \, , \label{gnuexact} 
\end{eqnarray}
where $\mathcal{L}_\nu(\xi)$ is the following function,
\begin{eqnarray}
	&& \hspace{-40pt} \mathcal{L}_\nu(\xi) = -\frac{i \nu T}{2 v_{\mathrm{F}}^2}  \int d\xi' \, U_{\mathrm{D}}(\xi') \left(x - x'\right) \nonumber \\
	&& \times \left[\cot \left(\pi T \left(z_\nu - z_\nu'\right) \right) + \cot \left(\pi T z_\nu'\right) \right] \, , \label{Lnu}
\end{eqnarray} 
where $z_\nu = \tau - i \nu x/v_{\mathrm{F}}$.
We used the symmetry property of the density-density interaction, $U_{\mathrm{D}}(-\xi) = U_{\mathrm{D}}(\xi)$.
We point out that the logarithmic part of the kernel $\mathcal{K}_{\mathrm{int}}(\xi)$, see the first part of Eq.~(\ref{Kint}), does not contribute to $g_{\nu \alpha} (\xi)$ at all.

We point out that Eq.~(\ref{gnuexact}) for the fermion Green function has been derived previously in Ref.~\cite{metznerFermiSystemsStrong1998} using the Ward identity approach applied in Ref.~\cite{dzyaloshinskiilarkin}. 
In order to compare Eq.~(\ref{gnuexact}) directly with Ref.~\cite{metznerFermiSystemsStrong1998}, we use Eqs.~(\ref{chiAAfinal}), (\ref{dMc}) to get the following representation,
\begin{eqnarray}
	&& \hspace{-29pt} g_{\nu \alpha}(\xi) = - \frac{\displaystyle T}{2 v_{\mathrm{F}} \displaystyle \sin\left(\pi T z_\nu\right)} \nonumber \\
	&& \hspace{-29pt} \times \exp\left[- T \sum\limits_{\omega_n} \int\limits_{-\infty}^\infty \frac{dq}{2 \pi}V(i \omega_n, q) \frac{e^{i q x - i \omega_n \tau} - 1}{\left(i \omega_n - \nu q v_{\mathrm{F}}\right)^2}\right] \, , \label{GMetzner}
\end{eqnarray}
which is exactly of the form of Eqs.~(5.46), (5.48) in Ref.~\cite{metznerFermiSystemsStrong1998}.
It has also been shown in Ref.~\cite{metznerFermiSystemsStrong1998} that non-singular RPA interactions $U_{\mathrm{D}}(\xi)$ do not destroy the Fermi liquid: the quasiparticle residue and the density of states at the Fermi level remain finite.
In this paper we mostly concentrate on non-singular finite-range interactions.
Authors of Ref.~\cite{metznerFermiSystemsStrong1998} also noticed that the double pole in Eq.~(\ref{GMetzner}) results in unphysical singularities near the mass shell.
Technical reason for these singularities comes from linearization of the electron spectrum near the Fermi level: the emission-absorption processes in the same-chirality channel are fully coherent, so the only source of the imaginary part of retarded self-energy is the vacuum polarization inbuilt in the RPA interaction.
This dangerous resonance results in the double-pole feature of $\mathcal{L}_\nu (i \omega_n, q)$ in Eq.~(\ref{GMetzner}) and leads to the mass-shell infrared catastrophe.
Natural physical resolution of this infrared catastrophe comes from the effects of spectral curvature [e.g. see Ref.~\cite{chubukov_effect_2006}] that we omitted during the bosonization procedure.
These spectral curvature effects are qualitatively important for correct resolution of the fermion spectral function very close to the mass shell even in 1D Luttinger liquids \cite{pustilnik_dynamic_2006,matveev_bosonization_2007,schmidt_spin-charge_2010}.
The spectral curvature effects on the emission-absorption processes are even more profound in higher dimensions \cite{chubukov_effect_2006}.
For this reason, Eq.~(\ref{GMetzner}) describes electron Green function away from the mass shell, and therefore, the Debye-Waller factor in Eq.~(\ref{GMetzner}) can only be produced by interaction-induced tail of spectral function.
In particular, if Eq.~(\ref{GMetzner}) provides an answer where the spectral function at $T = 0$ is still concentrated on the mass shell $\omega_\nu (q) = \nu q v_{\mathrm{F}}$, then such Debye-Waller factor can be fully subtracted.
As we will see below, this is what happens in case of a simplest model for a finite-range interaction.

Now, we assume that $U_{\mathrm{D}}(\xi)$ has a  finite range $R_{\mathrm{s}}  \gg 1/k_{\mathrm{F}}$.
As we are interested in the infrared regime when $|\tau| \gg 1/\Lambda$ and $|x| \gg v_{\mathrm{F}}/\Lambda$, where $\Lambda \sim v_{\mathrm{F}}/R_{\mathrm{s}}$ is the ultraviolet (UV) cutoff, then such interaction acts as effectively local kernel in Eq.~(\ref{Lnu}) leading to the following universal asymptotics,
\begin{eqnarray}
	&& \mathcal{L}_\nu (\xi) \approx -i \nu \gamma \frac{\pi T}{2 v_{\mathrm{F}}} x \cot \left(\pi T z_\nu\right) \, , \label{Lapprox}
\end{eqnarray}
where it is also assumed that $T \ll \Lambda$, and $\gamma$ is the dimensionless coupling constant,
\begin{eqnarray}
	&& \hspace{-20pt} \gamma = \int d\xi' \, \frac{U_{\mathrm{D}}(\xi')}{\pi v_{\mathrm{F}}} = \frac{2}{\pi v_{\mathrm{F}}} \int\limits_0^\infty U_{\mathrm{D}}(i \omega_n = 0, r) \, dr \, , \label{gamma}
\end{eqnarray}
where $U_{\mathrm{D}}(i \omega_n = 0, r)$ is the static component of the $D$-dimensional finite-range interaction.
A constant term in Eq.~(\ref{Lapprox}) is omitted,  other subleading corrections are suppressed by  $R_{\mathrm{s}}^2/|z_\nu|^2 \ll 1$ and therefore, they can be omitted in the infrared limit.
Now, we are ready to analyze the spectral function $\rho_{\nu\alpha}(t, x)$,
\begin{eqnarray}
	&& \hspace{-20pt} \rho_{\nu\alpha} (t, x) = \frac{g_{\nu \alpha}\left(i t - 0^+, x\right) - g_{\nu \alpha}\left(i t + 0^+, x\right)}{2 \pi} \, , \label{rhotx}
\end{eqnarray}
where we used analytic continuation of the Matsubara Green function via $\tau \to i t \pm 0^+$, $t$ is the real time.
As both $\mathcal{L}_\nu (\xi)$ and  $g_{\nu \alpha}(\xi)$ are analytic at $ \Lambda |z_{\nu}| \gg 1$, $\rho_{\nu\alpha} (t, x) = 0$ at this condition.
Therefore, the single-particle spectral function $\rho_{\nu\alpha}(t, x)$ is localized at $|v_{\mathrm{F}} t - \nu x| \lesssim R_{\mathrm{s}}$, i.e. there are no off-shell contributions.

We can show explicitly that $\rho_{\nu \alpha}(\omega, q)$ is fully localized at the mass shell.
For this, we evaluate the Green function $g_{\nu \alpha}(i \omega_n, q)$ at $T = 0$ explicitly, starting from Eq.~(\ref{Lapprox}) at $T \to 0$,
\begin{eqnarray}
	&& \hspace{-30pt} g_{\nu \alpha} (\tau, x) \approx  \frac{1}{2 \pi \left(i \nu x - v_{\mathrm{F}} \tau\right) } \exp\left[ \frac{\gamma}{2} \frac{i \nu x}{i \nu x - v_{\mathrm{F}} \tau}\right] \, . \label{gtauxspurious}
\end{eqnarray}
Expanding the exponential term in Taylor series and taking the Fourier transform, we find $g_{\nu \alpha}(i \omega, q)$,
\begin{eqnarray}
	&& \hspace{-20pt} g_{\nu \alpha}(i \omega, q) \approx \frac{1}{i \omega - \nu q v_{\mathrm{F}}} \exp\left(\frac{\gamma}{2} \frac{i \omega}{i \omega - \nu q v_{\mathrm{F}}}\right) \, . \label{gwqspurious}
\end{eqnarray}
The retarded Green function corresponds to the analytic continuation $i \omega \to \omega + i 0^+$, meaning that its imaginary part is zero at $\omega \ne \nu q v_{\mathrm{F}}$,
\begin{eqnarray}
	&& \hspace{-30pt} \rho_{\nu \alpha} (\omega, q) = - \frac{1}{\pi} \mathrm{Im}\left[g_{\nu \alpha} (\omega + i 0^+, q)\right] = 0, \hspace{3pt} \omega \ne \nu q v_{\mathrm{F}} \, .
\end{eqnarray}
The physical Green function is then equal to $g_{\nu \alpha}(i \omega, q)$ at $\nu = +1$ and $q = p - k_{\mathrm{F}}$ [distance from the Fermi surface], i.e. the physical spectral function is given by $\rho_{+1, \alpha}(\omega, q)$ which is zero away from the mass shell $\omega = q v_{\mathrm{F}}$.
For this reason, we consider the asymptotics given by Eqs.~(\ref{gtauxspurious}), (\ref{gwqspurious}) spurious: the only physical spectral function that is zero everywhere except one point $\omega = q v_{\mathrm{F}}$ must be the delta function.
Mathematically, infrared asymptotics of Euclidean correlators at $|x| \gg R_{\mathrm{s}}$, $v_{\mathrm{F}}|\tau| \gg R_{\mathrm{s}}$ implies $|z_\nu|^2 \gg R_{\mathrm{s}}^2$. 
In particular, after the analytic continuation $\tau \to i t \pm 0^+$, this condition requires $|x - v_{\mathrm{F}} t| \gg R_{\mathrm{s}}$. 
As the spectral function [see Eq.~(\ref{rhotx})] vanishes at this condition, this means that the asymptotic behavior given by Eq.~(\ref{gtauxspurious}) is spurious.

The spurious asymptotic form of Euclidean Green function given by Eq.~(\ref{gtauxspurious}) originates from the double-pole feature in $\mathcal{L}_\nu (i \omega_n, q)$, see Eq.~(\ref{GMetzner}). 
Subtracting the spurious asymptotics coming from the non-unitary double-pole feature in Eq.~(\ref{GMetzner}), we then find the following regularized asymptotic form of the Green function,
\begin{eqnarray}
	&& \hspace{-29pt} g_{\nu \alpha}(\xi) = - \frac{\displaystyle T}{2 v_{\mathrm{F}} \displaystyle \sin\left(\pi T z_\nu\right)}  \exp\left[T \sum\limits_{\omega_n} \int\limits_{-\infty}^\infty \frac{dq}{2 \pi}  \right. \nonumber \\
	&& \hspace{-29pt} \times
	\left. \frac{V(i \omega_n, q) - V(\nu q v_{\mathrm{F}}, q)}{\left(i \omega_n - \nu q v_{\mathrm{F}}\right)^2} \left(1 - e^{i q x - i \omega_n \tau}\right)\right] \, , \label{GMetznerreg}
\end{eqnarray}
where $z_\nu = \tau - i \nu x/v_{\mathrm{F}}$.
We point out that in true 1D system the unitarity of the Gaussian theory already demands $V(\nu q v_{\mathrm{F}}, q) = 0$.
However, such subtraction is required for the patch projection $V(i \omega_n, q)$ of the $D$-dimensional RPA interaction $U_{\mathrm{D}}(i \omega_n, p)$, see Eq.~(\ref{V1D}).
In next section, we introduce these counterterms in the interaction kernel $M_{\mathrm{int}}$, and derive its simplified expression for finite-range interactions.

As finite-range $D$-dimensional interactions do not alter single-particle spectral function, we provide a non-trivial example of dressed $D$-dimensional interactions causing the Luttinger-liquid scaling of the spectral function in Appendix~\ref{sec:LL}.
We point out that such interactions are singular, and take the form of gauge-boson mediated interactions.
This result qualitatively agrees with conclusions of Refs.~\cite{khveshchenkoLowenergyPropertiesTwodimensional1993,altshulerLowenergyPropertiesFermions1994,franz_mathrmqed_3_2002,franz_algebraic_2001}.

\section{Regularized bosonization procedure}
\label{sec:bosoreg}

In this section, we regularize bosonized expressions for two-point correlation functions derived in Sec.~\ref{sec:boso}. We also discuss the curvature effects that are neglected in bosonization.

First, we explicitly separate the double-pole features in the interaction kernel $M_{\mathrm{int}}(i \omega_n, q)$, see Eq.~(\ref{dMc}),
\begin{eqnarray}
	&& \hspace{-20pt} j_{\mathrm{A}, \mathrm{c}}^T M_{\mathrm{int}}(i \omega_n, q) j_{\mathrm{A}, \mathrm{c}} = \frac{\mathcal{J}_{\mathrm{A}}^2 - \mathcal{Q}_{\mathrm{A}}^2}{2} \frac{V(i \omega_n, q)}{\omega_n^2 + \left(q v_{\mathrm{F}}\right)^2} 
	\nonumber \\
	&& \hspace{-20pt}	+ \sum\limits_{\nu = \pm 1} \frac{\left(\mathcal{J}_{\mathrm{A}} + \nu \mathcal{Q}_{\mathrm{A}}\right)^2}{4} \frac{V(i \omega_n, q)}{(i \omega_n - \nu q v_{\mathrm{F}})^2} \, , \label{Mintchiralexpansion}
\end{eqnarray}
where $j_{\mathrm{A}, \mathrm{c}}$ is given by Eq.~(\ref{JAc}).
The first term in Eq.~(\ref{Mintchiralexpansion}) is already regular as it represents the mixed-chirality channel.
The sum over $\nu \in \{\pm 1\}$ in Eq.~(\ref{Mintchiralexpansion}) contains dangerous double poles that are responsible for spurious asymptotics of the Matsubara Green function that we discussed in Sec.~\ref{sec:GF}.
Now, it is clear that the single-particle Green function isolates single chiral sector as it corresponds to $\mathcal{J}_{\mathrm{A}} = \nu$, $\mathcal{Q}_{\mathrm{A}} = 1$.
As the multidimensional bosonization is unreliable near the mass shell, the Fourier transform of Eq.~(\ref{Mintchiralexpansion}) is not sensitive to analytic terms, as those do not contribute to the spectral function away from the mass shell. In fact, further dressing of the bare $1/z_{\nu}$ singularity of the Green function [see Eq.~(\ref{gnuexact}) at $T \to 0$] by higher powers of $1/z_{\nu}$ also does not produce any contribution to the spectral function away from the mass shell (they can be expressed in terms of derivatives of the delta function which are spurious mass-shell singularities that can be subtracted).
This brings the idea to regularize Eq.~(\ref{Mintchiralexpansion}) via subtracting out the non-unitary double-pole contribution and keeping the rest,
\begin{eqnarray}
	&& \hspace{-20pt} j_{\mathrm{A}, \mathrm{c}}^T M'_{\mathrm{int}}(i \omega_n, q) j_{\mathrm{A}, \mathrm{c}} = \frac{\mathcal{J}_{\mathrm{A}}^2 - \mathcal{Q}_{\mathrm{A}}^2}{2} \frac{V(i \omega_n, q)}{\omega_n^2 + \left(q v_{\mathrm{F}}\right)^2} 
	\nonumber \\
	&& \hspace{-20pt}	+ \sum\limits_{\nu = \pm 1} \frac{\left(\mathcal{J}_{\mathrm{A}} + \nu \mathcal{Q}_{\mathrm{A}}\right)^2}{4} \frac{V(i \omega_n, q) - V(\nu q v_{\mathrm{F}}, q)}{(i \omega_n - \nu q v_{\mathrm{F}})^2} \, . \label{Mintregularized}
\end{eqnarray}
We point out that the first term in Eq.~(\ref{Mintregularized}) does not require regularization as it is represented by the product of two simple poles.
We also emphasize that such regularization does not affect genuine 1D theory as unitarity of corresponding Gaussian theory requires $V(\nu q v_{\mathrm{F}}, q) = 0$.
This regularization is applied to the 1D patch projection $V(i \omega_n, q)$ of the $D$-dimensional RPA interaction, as in general such an operation does not preserve unitarity: non-analyticities in the spectral function can be negative, the unitarity is then restored by adding a smooth positive background.
Such subtraction is justified for non-singular interactions with finite limit $V(\nu q v_{\mathrm{F}}, q)$ at $q \to 0$.

The regularized interaction kernel is especially simple for finite-range interactions as those are determined by a single coupling constant $\gamma$, see Eq.~(\ref{gamma}),
\begin{eqnarray}
	&& \hspace{-20pt} j_{\mathrm{A}, \mathrm{c}}^T M'_{\mathrm{int}}(i \omega_n, q) j_{\mathrm{A}, \mathrm{c}} = \frac{\mathcal{J}_{\mathrm{A}}^2 - \mathcal{Q}_{\mathrm{A}}^2}{2} \frac{\pi v_{\mathrm{F}} \gamma}{\omega_n^2 + \left(q v_{\mathrm{F}}\right)^2}  \, , \label{MintfiniteR} \\
	&&  \hspace{-20pt} j_{\mathrm{A}, \mathrm{c}}^T M'_{\mathrm{int}}(\tau, x) j_{\mathrm{A}, \mathrm{c}} = \frac{\mathcal{Q}_{\mathrm{A}}^2 - \mathcal{J}_{\mathrm{A}}^2}{4} \gamma \ln\left|2 \sin \left(\pi T z\right)\right| \, , \label{Mintregtaux}
\end{eqnarray}
where the second equation is a 1D Fourier transform of the first one, and $z = \tau - i x/v_{\mathrm{F}}$.
The $T \to 0$ limit of Eq.~(\ref{Mintregtaux}) is proportional to $\ln|x^2 + (v_{\mathrm{F}} \tau)^2|$ which corresponds to the branch-cut singularity.
Such singularity strongly affects corresponding correlation function with $\mathcal{J}_{\mathrm{A}}^2 \ne \mathcal{Q}_{\mathrm{A}}^2$, and therefore results in power-law tails in corresponding spectral function.
Such off-shell contributions are accurately selected by the multidimensional bosonization, and therefore represent a physical non-analyticity.
For single-particle Green function $\mathcal{J}_{\mathrm{A}}^2 = \mathcal{Q}_{\mathrm{A}}^2 = 1$, hence this logarithmic term does not affect the Green function, which is consistent with our conclusion that single-particle Green function retains its Fermi-Liquid form.

Equation~(\ref{Mintregtaux}) is equivalent to the one-loop RG treatment of the finite-range interaction which has been done in Refs.~\cite{miserevMicroscopicMechanismPair2024,miserev_high-temperature_2025}. 
We point out that such one-loop RG implies that there is no running of corresponding coupling constant [see Ref.~\cite{miserevMicroscopicMechanismPair2024}] which also agrees with conclusions of Ref.~\cite{Shankar1994RGFermiLiquid}, where no running of the coupling constant was predicted in the full RG treatment of the finite-range forward-scattering interaction.
For finite-range interactions, this means that the running of the coupling constant is a higher order effect in $1/(k_{\mathrm{F}} R_{\mathrm{s}}) \ll 1$ originating from the finite-curvature effects as well as the gradient expansion with respect to the dressed forward-scattering interaction.
In multidimensional bosonization, these effects are treated as irrelevant.
However, if these effects determine the RG running of the coupling constant, then these effects are dangerously irrelevant, and therefore deserve a theoretical study on their own.
In this paper, we consider a standard multidimensional bosonization setting, where these effects are not taken into account.
We would also like to point out that other effects might contribute more to the RG scaling of the coupling constant than $1/(k_{\mathrm{F}} R_{\mathrm{s}})$ corrections to the multidimensional bosonization.
For example, this could be a short-range component of the interaction which cannot be treated via this approach, yet such interactions are often present in realistic physical systems [e.g., see Ref.~\cite{miserev_high-temperature_2025}].

Therefore, 1D two-point correlation functions corresponding to the finite-range interaction are given by Eq.~(\ref{chiAAfinal}) with $M_{\mathrm{int}}(\xi) \to M'_{\mathrm{int}}(\xi)$, where the latter is determined by Eq.~(\ref{Mintregtaux}).
We point out that $M'_{\mathrm{int}}(\tau \to 0, x \to 0)$ is not well defined.
However, we can fix it using the idea of a renormalization point: at $|z| \sim R_{\mathrm{s}}/v_{\mathrm{F}}$ correlation functions assume their non-interacting form.
Hence, $M'_{\mathrm{int}}(\tau \to 0, x \to 0)$ has to be interpreted as its value at the renormalization point, 
\begin{eqnarray}
	&& \hspace{-20pt} j_{\mathrm{A}, \mathrm{c}}^T \left[M'_{\mathrm{int}}(\tau, x) - M'_{\mathrm{int}}(0) \right] j_{\mathrm{A}, \mathrm{c}} \nonumber \\
	&& \hspace{-20pt} = \frac{\mathcal{Q}_{\mathrm{A}}^2 - \mathcal{J}_{\mathrm{A}}^2}{4} \gamma \ln\left|\frac{\Lambda}{\pi T}\sin \left(\pi T z\right)\right| \, , \label{Mintregtauxrenormpoint}
\end{eqnarray}
where $\Lambda = v_{\mathrm{F}}/R_{\mathrm{s}}$ is the UV cutoff, $|z| > 1/\Lambda$, $z = \tau - i x /v_{\mathrm{F}}$.
Here, we always assume that $T \ll \Lambda$, so $T$ represents a low-energy scale.

There is one more potential problem with Eq.~(\ref{Mintregtauxrenormpoint}): the long-distance asymptotics of Eq.~(\ref{Mintregtauxrenormpoint}) at $T > 0$ is linear at $|x| \gg R_{\mathrm{T}}$ meaning that at large enough value of $\gamma$ such a correlation function may diverge exponentially at $|x| \gg R_{\mathrm{T}}$, where we introduced the thermal length $R_{\mathrm{T}}$,
\begin{eqnarray}
	&& R_{\mathrm{T}} = \frac{v_{\mathrm{F}}}{2 \pi T} \, . \label{RT}
\end{eqnarray}
Such a behavior of a physical correlation function is not possible.
The problem comes from the contribution of the zero Matsubara frequency in Eq.~(\ref{MintfiniteR}) containing dangerous $\propto 1/q^2$ singularity at $ q \to 0$. 
This is similar to the double-pole issue we discussed before.
Close relation between multidimensional bosonization and RG suggests that the lowest energy scale this procedure can be trusted is $T$, meaning that the zero Matsubara frequency should be subtracted.
This can also be argued alternatively.
The multidimensional bosonization approach only takes into account non-analytic corrections to correlation functions.
Finite temperature $T > 0$ rounds up the non-analyticities, i.e. on the scale $|x| \gg R_{\mathrm{T}}$ correlation functions are analytic, hence this region is out of reach for the multidimensional bosonization.
Stopping this procedure at $|x| \sim R_{\mathrm{T}}$ is equivalent to subtracting the zero Matsubara frequency contribution.

This reasoning can be made explicit at the level of the building blocks of the correlators. At finite temperature, the non-interacting 1D Green function $g^{(0)}_\nu(\tau, x) = -T/[2 v_{\mathrm{F}} \sin(\pi T z_\nu)]$ [see Eq.~(\ref{g0taux})] decays exponentially at large spatial separation, $|g^{(0)}_\nu| \sim e^{-|x|/2 R_{\mathrm{T}}}$ for $|x| \gg R_{\mathrm{T}}$. Any two-point or composite correlator is a product of such propagators, and therefore inherits the same exponential envelope: at $|x| \gg R_{\mathrm{T}}$ it is exponentially small and analytic, so it cannot carry any non-analyticity. Consequently, the power-law singularities of the susceptibilities at $T \to 0$ cannot originate from the region $|x| \gg R_{\mathrm{T}}$; an unsubtracted contribution that grows there [as in Eq.~(\ref{Mintregtauxrenormpoint})] is necessarily spurious. This becomes quantitative in the Fourier transform below: after rescaling $r = R_{\mathrm{T}} u$ [see Eq.~(\ref{chiCST2kF})], the remaining dimensionless integral is convergent and dominated by $u \sim 1$, i.e. by distances $r \sim R_{\mathrm{T}}$, while the entire singular temperature dependence is carried by the prefactor $R_{\mathrm{T}}^{\gamma - \gamma_{\mathrm{c}}} \propto T^{-(\gamma - \gamma_{\mathrm{c}})}$ produced by the rescaling. The thermal length $R_{\mathrm{T}}$ thus plays a dual role: it is both the scale beyond which correlations are analytic --- and hence the natural point to terminate the bosonization, equivalently to subtract the zero Matsubara frequency --- and the scale that sets the power of $T$ governing the divergence of the susceptibility.

Therefore, the regularized expression for $M_{\mathrm{int}}(\xi)$ at $T > 0$ takes the following form,
\begin{eqnarray}
	&& \hspace{-20pt} j_{\mathrm{A}, \mathrm{c}}^T \left[M'_{\mathrm{int}}(\tau, x) - M'_{\mathrm{int}}(0) \right] j_{\mathrm{A}, \mathrm{c}} \nonumber \\
	&& \hspace{-20pt} = \frac{\mathcal{Q}_{\mathrm{A}}^2 - \mathcal{J}_{\mathrm{A}}^2}{4} \gamma \left[\ln\left|\frac{\Lambda}{\pi T}\sin \left(\pi T z\right)\right| - \frac{|x|}{2 R_{\mathrm{T}}} \right] \, . \label{MintregFINAL}
\end{eqnarray}
Therefore, 1D correlators considered in Sec.~\ref{sec:boso} [see Eq.~(\ref{chiAAfinal})] take the following form in case if the interaction is of a finite range,
\begin{eqnarray}
	&& \hspace{-20pt} \chi_{\mathrm{A} \mathrm{\overline{A}}}(\xi) = \chi^{(0)}_{\mathrm{A} \mathrm{\overline{A}}}(\xi) \left|\frac{\Lambda}{\pi T}\sin \left(\pi T z\right)\right|^{\frac{\displaystyle \gamma}{\displaystyle 4} \left(\mathcal{J}_{\mathrm{A}}^2 - \mathcal{Q}_{\mathrm{A}}^2\right)} \nonumber \\
	&& \hspace{13pt} \times \exp\left[-\frac{\gamma |x|}{8 R_{\mathrm{T}}} \left(\mathcal{J}_{\mathrm{A}}^2 - \mathcal{Q}_{\mathrm{A}}^2\right)\right] \, . \label{chiAAregularized}
\end{eqnarray}
Equation~(\ref{chiAAregularized}) is equivalent to the one-loop RG result derived for various two-particle susceptibilities in Refs.~\cite{miserevMicroscopicMechanismPair2024,miserev_high-temperature_2025}.
However, we stress that Eq.~(\ref{chiAAregularized}) is derived using the multidimensional bosonization, meaning that it corresponds to the full RG treatment: one-loop RG results presented in Refs.~\cite{miserevMicroscopicMechanismPair2024,miserev_high-temperature_2025} hold in all loops.
Moreover, Eq.~(\ref{chiAAregularized}) also allows us to calculate more complicated composite susceptibilities such as $2 N$-particle $2 N k_{\mathrm{F}}$ density wave susceptibilities as well as the $2 N$-particle $2 N e$ pair susceptibility.

There is one more important comment on the regularized expression for two-point correlators given by Eq.~(\ref{chiAAregularized}).
As we already discussed in Sec.~\ref{sec:GF}, the electron spectral function is zero away from the mass shell, hence true physical spectral function is determined by the dangerously irrelevant curvature and gradient expansion corrections omitted during the bosonization procedure.
They result in broadening of the spectral function over small infrared scale $\Lambda_{\mathrm{IR}} \ll \Lambda$.
If $T \gg \Lambda_{\mathrm{IR}}$, then subtraction of the zero Matsubara frequency in Eq.~(\ref{chiAAregularized}) safely removes dangerous sector near the mass shell of characteristic width $\Lambda_{\mathrm{IR}}$.
Therefore, Eq.~(\ref{chiAAregularized}) implies the following constraint on $T$,
\begin{eqnarray}
	&& \hspace{-20pt} \Lambda \gg T \gg \Lambda_{\mathrm{IR}} \, , \hspace{5pt} \Lambda = \frac{v_{\mathrm{F}}}{R_{\mathrm{s}}} \, , \hspace{5pt} \Lambda_{\mathrm{IR}} \sim \frac{1}{m R_{\mathrm{s}}^2} \sim \frac{\Lambda}{k_{\mathrm{F}} R_{\mathrm{s}}} \, , \label{LambdaIR}
\end{eqnarray}
where $m$ stands for the effective mass.
Here, we estimated $\Lambda_{\mathrm{IR}}$ as the scale attributed to the finite spectral curvature: the electron dispersion $\varepsilon(\bm q) = q_\parallel v_{\mathrm{F}} + \bm q^2/(2 m)$ [$q_\parallel$ is a 1D patch projection of $\bm q$, $|\bm q| \lesssim 1/R_{\mathrm{s}}$].
Therefore, the quadratic term in the dispersion yields a small but finite IR cut-off $\Lambda_{\mathrm{IR}} \sim \bm q^2/(2 m) \sim 1/(m R_{\mathrm{s}}^2)$ which we use in Eq.~(\ref{LambdaIR}).
Condition given by Eq.~(\ref{LambdaIR}) explicitly requires $k_{\mathrm{F}} R_{\mathrm{s}} \gg 1$.
If $k_{\mathrm{F}} R_{\mathrm{s}} \lesssim 1$, then $\Lambda \lesssim \Lambda_{\mathrm{IR}}$, leaving no finite interval for the multidimensional bosonization result given by Eq.~(\ref{chiAAregularized}).

\section{Charge, spin, and pair susceptibilities}
\label{sec:susc}

In this section we analyze charge, spin and pair susceptibilities given by Eqs.~(\ref{chiP1}), (\ref{chiC2kF}) using Eq.~(\ref{chiAAregularized}) for corresponding 1D susceptibilities derived for finite-range interactions.
At $T = 0$, results of this section agree with the one-loop RG treatment provided in Refs.~\cite{miserevMicroscopicMechanismPair2024,miserev_high-temperature_2025}.

\subsection{Pair susceptibility}

Let us start from the pair susceptibility given by Eq.~(\ref{chiP1}). The 1D pair susceptibility $\chi_{\mathrm{P}, \nu \nu'}^{(1D)}(\tau, r)$ is given by Eq.~(\ref{chiAAregularized}) with $\mathcal{Q}_{\mathrm{A}} = -2$, $\mathcal{J}_{\mathrm{A}}= -\nu - \nu'$. 
As $\mathcal{Q}_{\mathrm{A}}^2 = \mathcal{J}_{\mathrm{A}}^2$ at $\nu = \nu'$, the $2 k_{\mathrm{F}}$ pair susceptibility is not dressed by the finite-range interaction.
From now on, we concentrate on $\nu' = -\nu$ which corresponds to $\mathcal{J}_{\mathrm{A}} = 0$, i.e. it describes the pair susceptibility at zero momentum (Cooper susceptibility).
Then, we use Eq.~(\ref{chiAAregularized}) directly,
\begin{eqnarray}
	&& \hspace{-20pt} \chi_{\mathrm{P}, -\nu \nu}^{(1D)}(\xi) = \frac{\Lambda^{- \gamma}}{4 \pi^2 v_{\mathrm{F}}^2} \left|\frac{\pi T}{ \sin\left(\pi T z\right)} \right|^{2 + \gamma}  \exp\left[\frac{\gamma |x|}{2 R_{\mathrm{T}}}\right] \, . \label{chiPmnunu}
\end{eqnarray}
where $z = \tau - i x/v_{\mathrm{F}}$. 
Substituting Eq.~(\ref{chiPmnunu}) into Eq.~(\ref{chiPq0}), we find the Cooper pair susceptibility,
\begin{eqnarray}
	&& \hspace{-30pt} \chi_{\mathrm{P}}\left(\tau, r\right)\left.\right|_{p \sim 0} = \frac{\Lambda^2 \exp\left(\frac{\displaystyle \gamma r}{\displaystyle 2 R_{\mathrm{T}}}\right)}{2 \pi^2 v_{\mathrm{F}}^2 \left(\lambda_{\mathrm{F}} r\right)^{D - 1}} \left|\frac{\pi T}{\Lambda \sin \left(\pi T z\right)}\right|^{2 + \gamma}  , \label{chiPpowerlaw}
\end{eqnarray}
where $z = \tau - i r/v_{\mathrm{F}}$.
It is clear from Eq.~(\ref{chiPpowerlaw}) that attractive interactions with $\gamma < 0$ result in strong enhancement of the $p \sim 0$ pair susceptibility. 
The static Cooper pair susceptibility can be expressed in terms of a hypergeometric function,
\begin{eqnarray}
	&& \hspace{-25pt} \chi_{\mathrm{P}}\left(r\right)\left.\right|_{p \sim 0} = \frac{\left(2 \pi T \right)^{\gamma + 1}}{ \pi v_{\mathrm{F}}^2 \left(\lambda_{\mathrm{F}}r\right)^{D - 1} \Lambda^\gamma}  \exp\left(-\frac{r}{R_{\mathrm{T}}}\right) \nonumber \\ [5pt]
	&& \hspace{25pt} \times \,  {}_2F_1\left(1 + \frac{\gamma}{2}, 1 + \frac{\gamma}{2}; 1; e^{-2r/R_{\mathrm{T}}}\right) \, . \label{chiPr}
\end{eqnarray}
In order to derive Eq.~(\ref{chiPr}), we used the following identity,
\begin{eqnarray}
	&& \hspace{-20pt} \int\limits_0^\pi \frac{du}{\left|\sin(u + i s)\right|^{2 a}} = \pi 2^{2 a} e^{-2 a s} {}_2F_1\left(a,a;1;e^{-4 s}\right) \, , \label{sin2F1}
\end{eqnarray}
where $s > 0$ and $a$ is arbitrary.
Finally, the static Cooper pair susceptibility at $\omega_n = 0$ and $q = 0$ demonstrates the power-law behavior,
\begin{eqnarray}
	&& \hspace{-30pt} \chi_{\mathrm{P}}(T) = -\frac{N_{\mathrm{F}}}{N \gamma} \mathcal{A}_{\mathrm{P}}\left(\gamma\right) \left[ \left(\frac{2 \pi T}{\Lambda}\right)^\gamma - 1 \right] + \chi_{\mathrm{P}}\left(T_\Lambda\right) \, , \label{chiPT} \\
	&& \hspace{-30pt} \mathcal{A}_{\mathrm{P}}\left(\gamma\right) = -2 \gamma \int\limits_0^1 du \, \, {}_2F_1\left(1 + \frac{\gamma}{2}, 1 + \frac{\gamma}{2}; 1; u^2\right) \, , \label{AP}
\end{eqnarray}
where $N_{\mathrm{F}}/N$ is the density of states per flavor [see Eq.~(\ref{NF})], $T \ll \Lambda$, and $T_\Lambda$ is defined as follows:
\begin{eqnarray}
	&& T_\Lambda = \frac{\Lambda}{2 \pi} \, . \label{Tlambda}
\end{eqnarray}
Here, $\chi_{\mathrm{P}}\left(T_\Lambda\right) = \mathcal{O}(N_{\mathrm{F}}/N)$ corresponds to the UV renormalization point for $\chi_{\mathrm{P}}(T)$.
The function $\mathcal{A}_{\mathrm{P}}(\gamma)$ is well defined for $\gamma < 0$, strictly increasing with increasing $|\gamma|$, and approaches $1$ as $\gamma \to 0^-$.
The pair susceptibility is irrelevant at $\gamma > 0$, yet $\mathcal{A}_{\mathrm{P}}(\gamma)$ diverges at $\gamma > 0$. 
This divergence comes from the vicinity of $u = 1$ in Eq.~(\ref{AP}) which corresponds to short distances [from Eq.~(\ref{chiPr}), $u = e^{-r/R_{\mathrm{T}}}$].
Therefore, this divergence is not physical and is resolved by the UV cut-off at $r \sim R_{\mathrm{s}}$.
Therefore, we find that if $\gamma < 0$, $\chi_{\mathrm{P}}(T) \propto 1/T^{|\gamma|}$ diverges at $T \to 0$ as a power law, which is much faster than the logarithmic divergence of the particle-particle bubble.
We point out that Eq.~(\ref{chiPT}) coincides with the one-loop RG result derived in Refs.~\cite{miserevMicroscopicMechanismPair2024,miserev_high-temperature_2025} where $\mathcal{A}_{\mathrm{P}}(\gamma) \approx 1$ was used.

We point out that $\chi_{\mathrm{P}}(T)$ is finite at any finite $T$. In other words, there is no long-range order if the interaction is of the forward-scattering type due to quasi-1D physics of many-body correlations.
In Ref.~\cite{miserev_high-temperature_2025}, we demonstrated that the long-range superconducting order is stabilized by a short-range attractive interaction, where the forward-scattering pair susceptibility $\chi_{\mathrm{P}}(T)$ replaces the BCS particle-hole bubble. 
In Ref.~\cite{miserev_high-temperature_2025}, we performed the one-loop RG and therefore considered $\mathcal{A}_{\mathrm{P}}(\gamma) \approx 1$ which is valid at $|\gamma| \ll 1$.
In this work, we demonstrated that the one-loop RG result for $\chi_{\mathrm{P}}(T)$ holds in all loops, hence Eq.~(\ref{chiPT}) provides the leading contribution coming from the forward-scattering component of the finite-range interaction at any coupling.

\subsection{Charge and spin susceptibilities}

Now, we calculate charge and spin susceptibilities given by Eq.~(\ref{chiC2kF}).
The 1D vertex operator is of the form $A(\xi) \propto \overline{\psi}_{\nu \alpha}(\xi) \psi_{-\nu \alpha'}(\xi)$ corresponding to $\mathcal{Q}_{\mathrm{A}} = 0$ and $\mathcal{J}_{\mathrm{A}} = -2\nu$.
Substituting this in Eq.~(\ref{chiAAregularized}), we find 1D $2k_{\mathrm{F}}$ spin/charge susceptibility,
\begin{eqnarray}
	&& \hspace{-30pt} \chi_{\mathrm{C/S}, -\nu \nu}^{(1D)}(\xi) = \frac{\Lambda^{\gamma}}{4 \pi^2 v_{\mathrm{F}}^2} \left|\frac{\pi T}{ \sin\left(\pi T z\right)} \right|^{2 - \gamma} \!\! \exp\left[-\frac{\gamma |x|}{2 R_{\mathrm{T}}}\right] \, . \label{chiCS1D2kF}
\end{eqnarray}
This expression differs from Eq.~(\ref{chiPmnunu}) only by a substitution $\gamma \to - \gamma$.
The static 1D susceptibility then follows from Eq.~(\ref{sin2F1}),
\begin{eqnarray}
	&&  \hspace{-30pt} \chi_{\mathrm{C/S}, -\nu \nu}^{(1D)}(r) = \left(\frac{\Lambda}{2 \pi T}\right)^{\gamma} \frac{T}{v_{\mathrm{F}}^2} \exp\left(-\frac{r}{R_{\mathrm{T}}}\right) \nonumber \\ [5pt]
	&&  \hspace{25pt} \times \, \, {}_2F_{1} \left(1-\frac{\gamma}{2}, 1 - \frac{\gamma}{2}; 1; e^{-2 r/R_{\mathrm{T}}}\right) \, , \label{chiCS1Dr}
\end{eqnarray}
where $r = |x|$, the susceptibility is evaluated at $\omega_n = 0$.
Substituting this into Eq.~(\ref{chiC2kF}) evaluated at zero Matsubara frequency, we find static $2k_F$ spin/charge susceptibility,
\begin{eqnarray}
	&& \hspace{-30pt} \chi_{\mathrm{C/S}}\left(r\right)\left.\right|_{2 k_{\mathrm{F}}} = \left(\frac{\Lambda}{2 \pi T}\right)^{\gamma} \frac{\exp\left(-\frac{\displaystyle r}{\displaystyle R_{\mathrm{T}}} \right)}{\pi v_{\mathrm{F}} R_{\mathrm{T}}} \nonumber \\ [5pt] 
	&& \hspace{-30pt} \times \frac{\cos\left(2 k_{\mathrm{F}} r - 2 \vartheta_{\mathrm{D}}\right)}{\left(\lambda_{\mathrm{F}} r\right)^{D - 1}} \, {}_2F_{1} \left(1-\frac{\gamma}{2}, 1 - \frac{\gamma}{2}; 1; e^{-2 r/R_{\mathrm{T}}}\right) \, . \label{chiCS2kFrD}
\end{eqnarray}
Next, we evaluate $\chi_{\mathrm{C/S}}\left(T\right)$ by taking the $D$-dimensional Fourier transform at momentum $\bm p$ with $p = |\bm p| = 2 k_{\mathrm{F}}$,
\begin{eqnarray}
	&& \chi_{\mathrm{C/S}}\left(T\right) = \int d\bm r \, e^{-i \bm p \cdot \bm r} \chi_{\mathrm{C/S}}\left(r\right)\left.\right|_{2 k_{\mathrm{F}}} \, . \label{chiCSTdef}
\end{eqnarray}
Substituting the angular integral given by Eq.~(\ref{angleint}) into Eq.~(\ref{chiCSTdef}), we find,
\begin{eqnarray}
	&& \chi_{\mathrm{C/S}}\left(T\right) = 2 \pi \left(\frac{\lambda_{\mathrm{F}}}{2}\right)^{\frac{D}{2}-1} \nonumber \\
	&& \times \int\limits_0^\infty dr\, r^{\frac{D}{2}} J_{\frac{D}{2}-1}(2 k_{\mathrm{F}} r) \chi_{\mathrm{C/S}}\left(r\right)\left.\right|_{2 k_{\mathrm{F}}} \, , \label{chiCST1}
\end{eqnarray}
where $\lambda_{F} = 2 \pi/k_{\mathrm{F}}$ is the Fermi wavelength, and we used $p = 2 k_{\mathrm{F}}$.
The integral in Eq.~(\ref{chiCST1}) converges at $r \to 0$, so no UV cut-off is required.
We can significantly simplify Eq.~(\ref{chiCST1}) by noticing that the integral over $r$ converges at large distances $r \sim R_{\mathrm{T}} \gg 1/k_{\mathrm{F}}$, so we can use the large-argument asymptotics of Bessel function,
\begin{eqnarray}
	&& J_{\frac{D}{2}-1}(p r) \approx \sqrt{\frac{2}{\pi p r}} \cos \left(p r - \vartheta_{\mathrm{D}}\right) \, , \label{Besselasympt}
\end{eqnarray}
where $p r \gg 1$, and $\vartheta_{\mathrm{D}}$ is given by Eq.~(\ref{thetaphase}).
Substituting Eqs.~(\ref{chiCS2kFrD}), (\ref{Besselasympt}) into Eq.~(\ref{chiCST1}), we find that the integrand contains the product of two quickly oscillating cosines that can be rewritten as a sum of ``fast'' cosine term with $\sim 4 k_{\mathrm{F}} r$ argument and ``slow'' cosine with $\sim (p - 2 k_{\mathrm{F}}) r$ argument.
The ``fast'' term converges quickly on the scale $r \sim 1/k_{\mathrm{F}}$, and it cannot even be included as Eq.~(\ref{Besselasympt}) implies that $r \gg 1/k_{\mathrm{F}}$.
The second cosine becomes constant at $p = 2k_{\mathrm{F}}$.
Taking this into account and rescaling the coordinate $r = R_{\mathrm{T}} u$, we get the following expression,
\begin{eqnarray}
	&& \hspace{-28pt} \chi_{\mathrm{C/S}}\left(T\right) = \frac{\mathcal{A}_{\mathrm{C/S}}\left(\gamma\right)}{\pi v_{\mathrm{F}}}  \frac{\cos\vartheta_{\mathrm{D}} \, \Gamma\left(1 - \gamma_{\mathrm{c}}\right)}{\left(2 \lambda_{\mathrm{F}} R_{\mathrm{s}}\right)^{\gamma_{\mathrm{c}}}} \nonumber \\
	&& \hspace{12pt} \times \left[\left(\frac{\Lambda}{2\pi T}\right)^{\gamma - \gamma_{\mathrm{c}}} - 1\right] + \chi_{\mathrm{C/S}}\left(T_{\mathrm{\Lambda}}\right) \, , \label{chiCST2kF} \\
	&& \hspace{-28pt} \mathcal{A}_{\mathrm{C/S}}\left(\gamma\right) = \int\limits_0^\infty \frac{du \,  e^{-u} u^{-\gamma_{\mathrm{c}}}}{\Gamma\left(1 - \gamma_{\mathrm{c}}\right)} \, \nonumber \\
	&& \hspace{12pt} \times  {}_2F_1\left(1-\frac{\gamma}{2},1 - \frac{\gamma}{2}; 1; e^{-2 u}\right) \, , \\
	&& \hspace{-30pt} \gamma_{\mathrm{c}} = \frac{D - 1}{2} \, , \label{gammac}
\end{eqnarray}
where $\chi_{\mathrm{C/S}}\left(T_{\mathrm{\Lambda}}\right) = \mathcal{O}(N_{\mathrm{F}}/N)$.
The susceptibility is relevant at $\gamma > \gamma_{\mathrm{c}}$.
We point out that the integral representation for $\mathcal{A}_{\mathrm{C/S}}\left(\gamma\right)$ at $\gamma > \gamma_{\mathrm{c}}$ is well behaved.  
The following series representation might also be useful for practical purposes,
\begin{eqnarray}
	&& \hspace{-30pt} \mathcal{A}_{\mathrm{C/S}}\left(\gamma\right) = \sum\limits_{n = 0}^\infty \left[\frac{\Gamma\left(1 - \frac{\gamma}{2} + n\right)}{\Gamma\left(1 - \frac{\gamma}{2}\right) n!}\right]^2 \frac{1}{\left(2 n + 1\right)^{1 - \gamma_{\mathrm{c}}}} \, , \label{ACSseries}
\end{eqnarray}
where the sum is convergent at $\gamma > \gamma_{\mathrm{c}}$.
The scaling with $T$ agrees with the one-loop RG results in Ref.~\cite{miserevMicroscopicMechanismPair2024}.
If $\gamma \to \gamma_{\mathrm{c}}$, $\mathcal{A}_{\mathrm{C/S}}(\gamma) \propto 1/(\gamma - \gamma_{\mathrm{c}})$, so at $\gamma = \gamma_{\mathrm{c}}$ the susceptibility is logarithmic, $\chi_{\mathrm{C/S}}(T) \propto \ln[\Lambda/(2 \pi T)]$, as expected at the quantum critical point.

The condition $\gamma > \gamma_{\mathrm{c}}$ ensures that the susceptibility is growing at decreasing $T$. However, the multidimensional bosonization cannot be trusted all the way down to $T = 0$ due to the infrared cutoff $\Lambda_{\mathrm{IR}}$ introduced in Eq.~(\ref{LambdaIR}).
Therefore, maximal enhancement of $\chi_{\mathrm{C/S}}(T)$ happens at $T_{\mathrm{IR}} \sim \Lambda_{\mathrm{IR}} \sim \Lambda/(k_{\mathrm{F}}R_{\mathrm{s}})$ such that the maximal non-analytic enhancement of the susceptibility can be estimated as follows,
\begin{eqnarray}
	&& \hspace{-10pt} \chi_{\mathrm{C/S}}\left(T_{\mathrm{IR}}\right) - \chi_{\mathrm{C/S}}\left(T_{\mathrm{\Lambda}}\right) \propto \frac{N_{\mathrm{F}}}{N} \left(k_{\mathrm{F}} R_{\mathrm{s}}\right)^{\gamma - 2 \gamma_{\mathrm{c}}} \, ,
\end{eqnarray}
where the proportionality coefficient depends on $\gamma$ and $D$ but generally is order one.
Additional $(k_{\mathrm{F}} R_{\mathrm{s}})^{-\gamma_{\mathrm{c}}}$ comes from the prefactor in Eq.~(\ref{chiCST2kF}).
As $\chi_{\mathrm{C/S}}\left(T_{\mathrm{\Lambda}}\right) = \mathcal{O}(N_{\mathrm{F}}/N)$, this non-analytic correction is dominant only at $\gamma > 2 \gamma_{\mathrm{c}}$.

Similarly to $\chi_{\mathrm{P}}(T)$,  $\chi_{\mathrm{C/S}}(T)$ is finite at $T > 0$, i.e. no long-range charge/spin density-wave order is possible. 
In order to push the system towards ordering, short-range interaction that is present in any realistic electron system, must be included.
However, this topic contains some more technical details and new physical results, and therefore deserves a separate publication.
Results of this work can be used as a launch platform for studying charge/spin density-wave orders in presence of strong forward-scattering and weak short-range repulsive or attractive interactions, where the latter can be taken into account via the RG approach.

\section{Composite susceptibilities}
\label{sec:higherorder}

In this section, we consider composite susceptibilities with the vertex operator $A(\xi)$ [see Eq.~(\ref{A})] which is a product of more than two fermionic operators.
To the best of our knowledge, results of this section are original.

First, fix total number $N_{\mathrm{A}} = n + m$ of the fermion operators in $A(\xi)$ defined in Eq.~(\ref{A}).
Given choice of $A(\xi)$ corresponds to the following harmonic of corresponding $D$-dimensional susceptibility which we denote here as $\chi^{\mathrm{(D)}}_{\mathrm{A \overline{A}}} (\tau, r)$,
\begin{eqnarray}
	&& \chi^{\mathrm{(D)}}_{\mathrm{A \overline{A}}} (\tau, r) = \frac{e^{i \mathcal{J}_{\mathrm{A}} \left(k_{\mathrm{F}} r - \vartheta_{\mathrm{D}}\right)}}{\left(\lambda_{\mathrm{F}} r\right)^{\frac{N_{\mathrm{A}}}{2}(D - 1)}} \chi_{\mathrm{A \overline{A}}} (\tau, r) \, , \label{chiAAD}
\end{eqnarray} 
where $\chi_{\mathrm{A \overline{A}}} (\tau, r)$ is the 1D susceptibility defined in Eq.~(\ref{chiAA}), $\mathcal{J}_{\mathrm{A}}$ is the total ``current'' carried by the vertex operator [see Eq.~(\ref{QJ})].
Full $D$-dimensional susceptibility is represented by the sum over all its harmonics.
However, here we study them separately in order to identify the most relevant harmonics.
According to Eq.~(\ref{chiAAregularized}), the most relevant susceptibility for attractive (repulsive) interactions corresponds to the minimum (maximum) of $\mathcal{J}_{\mathrm{A}}^2 - \mathcal{Q}_{\mathrm{A}}^2$,
\begin{equation}
	\left\{
	\begin{array}{cc}
		\left|\mathcal{Q}_{\mathrm{A}}\right| = N_{\mathrm{A}} \, , \hspace{5pt} \left|\mathcal{J}_{\mathrm{A}}\right| = N_{\mathrm{A}} \,  \mathrm{mod} \, 2 \, , \hspace{5pt} & \gamma < 0 \\
		\left|\mathcal{Q}_{\mathrm{A}}\right| = N_{\mathrm{A}} \, \mathrm{mod} \, 2 \, , \hspace{5pt} \left|\mathcal{J}_{\mathrm{A}}\right| = N_{\mathrm{A}} \, , \hspace{5pt} & \gamma > 0 \\
	\end{array}
	\right. \label{QJA}
\end{equation}
where $N_{\mathrm{A}} \,  \mathrm{mod} \, 2 \, \in \{0, 1\}$ is the residue of $N_{\mathrm{A}}$ modulo 2.

Here, we consider bosonic correlators corresponding to the composite susceptibilities, i.e. $N_{\mathrm{A}} = 2 \mathcal{N}_{\mathrm{A}}$ is even number.
Fermionic composite correlators can be evaluated following similar procedure. 
We point out that such fermionic correlators can be utilized for evaluation of the anomalous self-energy in case if corresponding condensate is present.
In this paper, we are only interested in conditions for different instabilities at infinitesimal condensate or near the critical temperature.
Anomalous self-energies in presence of a finite condensate are a matter of future study.

\subsection{Composite pair susceptibility}

For attractive finite-range interactions, $\gamma < 0$, the most relevant susceptibility with fixed $N_{\mathrm{A}}$ corresponds to the zero-momentum pair susceptibility probing the instability towards $N_{\mathrm{A}} e$ superconducting state, where $e$ stands for the elementary charge,
\begin{eqnarray}
	&& \hspace{-51pt} \chi_{\mathrm{P,N_{\mathrm{A}}}}(\tau, r) = \frac{\Lambda^{-\gamma \mathcal{N}_{\mathrm{A}}^2} \exp\left(\mathcal{N}_{\mathrm{A}}^2 \frac{\gamma r}{2 R_{\mathrm{T}}}\right)}{\left(2 \pi v_{\mathrm{F}}\right)^{2 \mathcal{N}_{\mathrm{A}}}\left(\lambda_{\mathrm{F}} r\right)^{\mathcal{N}_{\mathrm{A}}(D - 1)}} \nonumber \\
	&& \times \left|\frac{\pi T}{\sin\left(\pi T z\right)} \right|^{2 \mathcal{N}_{\mathrm{A}} + \gamma \mathcal{N}_{\mathrm{A}}^2} \hspace{-2pt} , 
\end{eqnarray}
where $z = \tau - i r/v_{\mathrm{F}}$, and $\mathcal{N}_{\mathrm{A}}$ corresponds to number of pairs forming a single $2 \mathcal{N}_{\mathrm{A}} e$ charge ``unit''.
We introduced a separate notation $\chi_{\mathrm{P,N_{\mathrm{A}}}}(\tau, r)$ for the $D$-dimensional multi-pair susceptibility in order to avoid confusion with other susceptibilities that we consider in this section.
Using Eq.~(\ref{sin2F1}) to evaluate the susceptibility on zero Matsubara frequency, we find,
\begin{eqnarray}
	&& \hspace{-30pt} \chi_{\mathrm{P,N_{\mathrm{A}}}}(r) = \left(\frac{2 \pi}{\Lambda}\right)^{\gamma \mathcal{N}_{\mathrm{A}}^2} \frac{T^{\alpha_{\mathrm{P}} - 1}}{v_{\mathrm{F}}^{2 \mathcal{N}_{\mathrm{A}}}} \frac{\exp\left(-\mathcal{N}_{\mathrm{A}} \frac{\displaystyle r}{\displaystyle R_{\mathrm{T}}}\right)}{\left(\lambda_{\mathrm{F}} r\right)^{\mathcal{N}_{\mathrm{A}} (D - 1)}} \nonumber \\ [5pt]
	&& \hspace{10pt} \times {}_2F_1\left(\frac{\alpha_{\mathrm{P}}}{2},\frac{\alpha_{\mathrm{P}}}{2}; 1; e^{-2 r/R_{\mathrm{T}}} \right) \, , \label{chiPNAr} \\ [5pt]
	&& \hspace{-30pt} \alpha_{\mathrm{P}} = 2 \mathcal{N}_{\mathrm{A}} + \gamma \mathcal{N}_{\mathrm{A}}^2 \, . \label{alphaP}
\end{eqnarray}
In case of $\mathcal{N}_{\mathrm{A}} = 1$, we recover Eq.~(\ref{chiPr}) up to the factor of 2, which comes from two equivalent chirality configurations in Eq.~(\ref{chiPq0}).
Equation~(\ref{chiPNAr}) is written for a single choice of the chirality indices. 
Full susceptibility must be multiplied by corresponding combinatorial factor counting all equivalent choices of the chirality indices.
The $D$-dimensional zero-momentum Fourier transform of Eq.~(\ref{chiPNAr}) is the following,
\begin{eqnarray}
	&& \hspace{-20pt} \chi_{\mathrm{P,N_{\mathrm{A}}}}\left(T\right) = \chi_{\mathrm{P,N_{\mathrm{A}}}}\left(T_{\mathrm{\Lambda}}\right) + \mathcal{A}_{\mathrm{P}, \mathrm{\mathcal{N}}_{\mathrm{A}}} \left(\alpha_{\mathrm{P}}\right)	 \frac{N_{\mathrm{F}}}{N}
	k_{\mathrm{F}}^{2 D \left(\mathcal{N}_{\mathrm{A}}- 1\right)}  
	 \nonumber \\ [5pt]
	&& \hspace{0pt} \times  \frac{ \left(T/T_{\mathrm{\Lambda}} \right)^{\gamma \mathcal{N}_{\mathrm{A}}^2 + \left(\mathcal{N}_{\mathrm{A}} - 1\right) \left(D + 1\right)} - 1}{\left(2 \pi k_{\mathrm{F}} R_{\mathrm{s}}\right)^{\left(\mathcal{N}_{\mathrm{A}} - 1\right) \left(D + 1\right)} } \, , \label{chiPNAT} \\ [5pt]
	&& \hspace{-20pt} \mathcal{A}_{\mathrm{P}, \mathrm{\mathcal{N}}_{\mathrm{A}}} \left(\alpha_{\mathrm{P}}\right) \nonumber \\ [5pt]
	&& \hspace{0pt}  = \int\limits_0^\infty \frac{du \, e^{-\mathcal{N}_{\mathrm{A}} u}}{u^{(\mathcal{N}_{\mathrm{A}} - 1)(D - 1)}} \, {}_2F_1\left(\frac{\alpha_{\mathrm{P}}}{2}, \frac{\alpha_{\mathrm{P}}}{2}; 1; e^{-2 u} \right) \, , \label{APNA}
\end{eqnarray}
where $T_{\mathrm{\Lambda}}$ is the UV cutoff temperature, see Eq.~(\ref{Tlambda}).
Here, the constant part $\chi_{\mathrm{P,N_{\mathrm{A}}}}\left(T_{\mathrm{\Lambda}}\right) = \mathcal{O}(N_{\mathrm{F}} 	k_{\mathrm{F}}^{2 D \left(\mathcal{N}_{\mathrm{A}}- 1\right)}/N)$ is the renormalization point at large temperature $T = T_{\mathrm{\Lambda}}$ [see Eq.~(\ref{Tlambda})].
When the $\mathcal{N}_{\mathrm{A}}$-pair susceptibility $\chi_{\mathrm{P,N_{\mathrm{A}}}}\left(T\right)$ diverges at $T \to 0$, the integral $\mathcal{A}_{\mathrm{P}, \mathrm{\mathcal{N}}_{\mathrm{A}}} \left(\alpha_{\mathrm{P}}\right)$ is convergent at $u \to 0$, so no ultraviolet regularization is required in this case.
The susceptibility $\chi_{\mathrm{P,N_{\mathrm{A}}}}\left(T\right)$ is relevant when negative $\gamma$ is below the following critical value,
\begin{eqnarray}
	&& |\gamma| > \gamma\left(\mathcal{N}_{\mathrm{A}}\right) = \frac{\left(\mathcal{N}_{\mathrm{A}} - 1\right) \left(D + 1\right)}{\mathcal{N}_{\mathrm{A}}^2} \, . \label{gammaNA}
\end{eqnarray}
Of course, the usual pair susceptibility is relevant at any $\gamma < 0$.
Now, we consider the competition between composite pair susceptibilities with $\mathcal{N}_{\mathrm{A}} \ge 2$.
As $\gamma(\mathcal{N}_{\mathrm{A}})$ is an increasing function of $\mathcal{N}_{\mathrm{A}}$ at $\mathcal{N}_{\mathrm{A}} \ge 2$, the composite pair susceptibility with largest possible $\mathcal{N}_{\mathrm{A}} \ge 2$ becomes relevant first, and remains the most relevant amid all composite susceptibilities for more negative values of $\gamma$ [the latter is guaranteed by steeper $\gamma$-slope in the $T$-exponent in Eq.~(\ref{chiPNAT})].
Therefore, it only makes sense to consider the largest possible $\mathcal{N}_{\mathrm{A}} = N$, so $N_{\mathrm{A}} = 2N$.
The interesting situation occurs when $\chi_{\mathrm{P,2N}}\left(T\right)$ becomes more relevant than usual pair susceptibility $\chi_{\mathrm{P}}\left(T\right)$, see Eq.~(\ref{chiPT}), which happens at the following negative $\gamma < 0$,
\begin{eqnarray}
	&& \left|\gamma \right| > \gamma_{\mathrm{2N}} \equiv \frac{D + 1}{N + 1} \, . \label{gammaN}
\end{eqnarray}
In other words, we expect normal $2e$ superconductivity at $ -\gamma_{\mathrm{2N}} < \gamma < 0$, and $2 N e$ superconductivity at $\gamma < -\gamma_{\mathrm{2N}}$.
For example, if $N = 2$ (spin degeneracy), the $4e$ superconductivity is possible if $\gamma < -(D+1)/3$.

However, the relevance condition given by Eq.~(\ref{gammaNA}) is not enough to ensure the dominance of calculated non-analytic enhancement compared to the non-interacting background value $\chi_{\mathrm{P,2 N}}\left(T_{\mathrm{\Lambda}}\right)$ due to the infrared temperature cutoff, see Eq.~(\ref{LambdaIR}).
The non-analytic part of Eq.~(\ref{chiPNAT}) [we consider $\mathcal{N}_{\mathrm{A}} = N$] at lowest temperature $T_{\mathrm{IR}} \sim \Lambda_{\mathrm{IR}}$ is parametrically greater than $\chi_{\mathrm{P,2 N}}\left(T_{\mathrm{\Lambda}}\right)$ only if the following condition holds,
\begin{eqnarray}
	&& |\gamma| > 2 \gamma(N) \, , \label{dominancePN}
\end{eqnarray}
where $\gamma(N)$ is introduced in Eq.~(\ref{gammaNA}).
As for composite pair susceptibility $N \ge 2$, then $2 \gamma(N) > \gamma_{\mathrm{2 N}}$, i.e. if $\chi_{\mathrm{P,2 N}}\left(T\right)$ is dominant [i.e. Eq.~(\ref{dominancePN}) is satisfied], then it is automatically more relevant than the usual pair susceptibility. 
For this reason, we consider Eq.~(\ref{dominancePN}) as a necessary condition for $2 N e$ superconductivity instead of weaker relevance constraint $|\gamma| > \gamma(N)$, see Eq.~(\ref{gammaNA}).
For example, if $N = 2$, the $4e$ superconductivity is expected at $|\gamma| > (D + 1)/2$.

\subsection{Composite density wave susceptibilities}

The situation is similar in case of repulsive interactions [$\gamma > 0$], where relevant bosonic susceptibilities measure an instability towards charge/spin density waves.
In full analogy to the attractive case, only $N_{\mathrm{A}} = 2 N$ composite density-wave susceptibility can become the most relevant, this happens when $\gamma > \gamma_{\mathrm{2N}}$, meanwhile the $2k_{\mathrm{F}}$ susceptibility is the most relevant at $\gamma_{\mathrm{2 N}} > \gamma > (D - 1)/2$, where $\gamma_{\mathrm{2N}}$ is given by Eq.~(\ref{gammaN}).
The $2Nk_{\mathrm{F}}$ charge/spin susceptibility is given by the following expression,
\begin{eqnarray}
	&& \hspace{-25pt} \chi_{\mathrm{C/S,2N}}(\tau, r) = \frac{\Lambda^{\gamma N^2}}{\left(2 \pi v_{\mathrm{F}}\right)^{2 N}} \frac{2 \cos\left[2N \left(k_{\mathrm{F}} r - \vartheta_{\mathrm{D}}\right)\right]}{\left(\lambda_{\mathrm{F}} r\right)^{N(D - 1)}} \nonumber \\
	&& \times \left|\frac{\pi T}{\sin\left(\pi T z\right)} \right|^{2 N - \gamma N^2} \exp\left(-N^2 \frac{\gamma r}{2 R_{\mathrm{T}}}\right)  \, , \label{chiCS2Ntaur}
\end{eqnarray}
where $z = \tau -i r/v_{\mathrm{F}}$.
Using Eq.~(\ref{sin2F1}), we find the static $2Nk_{\mathrm{F}}$ charge/spin susceptibility,
\begin{eqnarray}
	&& \hspace{-45pt} \chi_{\mathrm{C/S,2N}}(r) = \left(\frac{\Lambda}{2 \pi}\right)^{\gamma N^2} \frac{T^{\alpha_{\mathrm{C/S}}-1}}{v_{\mathrm{F}}^{2N}}  \nonumber \\ [5pt]
	&& \hspace{0pt} \times \frac{2 \cos\left[2N \left(k_{\mathrm{F}} r - \vartheta_{\mathrm{D}}\right)\right]}{\left(\lambda_{\mathrm{F}} r\right)^{N(D - 1)}}  \exp\left(-\frac{N r}{R_{\mathrm{T}}} \right) \nonumber \\ [5pt]
	&& \hspace{0pt} \times {}_2F_1\left(\frac{\alpha_{\mathrm{C/S}}}{2}, \frac{\alpha_{\mathrm{C/S}}}{2}; 1; e^{-2r/R_{\mathrm{T}}} \right) \, , \label{chiCS2Nr} \\ [5pt]
	&& \hspace{-45pt} \alpha_{\mathrm{C/S}} = 2 N - \gamma N^2 \, . \label{alphaCS}
\end{eqnarray}
At $N = 1$, Eq.~(\ref{chiCS2Nr}) coincides with Eq.~(\ref{chiCS2kFrD}) for the static $2 k_{\mathrm{F}}$ charge/spin susceptibility.
The $D$-dimensional Fourier transform is calculated following the same recipe as for the $2k_{\mathrm{F}}$ charge/spin susceptibility,
\begin{eqnarray}
	&& \hspace{-30pt} \chi_{\mathrm{C/S, 2N}}(T)\left.\right|_{2Nk_{\mathrm{F}}} = \frac{\cos\left[\left(2 N - 1\right) \vartheta_{\mathrm{D}}\right]}{\pi (2N)^{\frac{D - 1}{2}} v_{\mathrm{F}}^{2N - 1}} \mathcal{A}_{\mathrm{C/S, 2N}}\left(\gamma\right)  \nonumber \\ [5pt]
	&& \hspace{-30pt} \times \left(\frac{k_{\mathrm{F}}}{v_{\mathrm{F}}}\right)^{\left(N - \frac{1}{2}\right)(D-1)} T_\Lambda^{\gamma N^2}  T^{(N - 1)(D + 1) + \frac{D - 1}{2} - \gamma N^2} \, , \label{chiCS2NT} \\ [5pt]
	&& \hspace{-30pt} \mathcal{A}_{\mathrm{C/S, 2N}}\left(\gamma\right) = \int\limits_0^\infty du\, u^{(1-2 N)\frac{D-1}{2}} e^{-N u} \nonumber \\
	&& \hspace{22pt} \times {}_2F_1\left(\frac{\alpha_{\mathrm{C/S}}}{2}, \frac{\alpha_{\mathrm{C/S}}}{2}; 1; e^{-2 u}\right) \, ,
\end{eqnarray}
where $T_\Lambda$ is given by Eq.~(\ref{Tlambda}).
Here, we only presented the non-analytic $T$-dependent part; a constant term taken at the renormalization point $T = T_{\mathrm{\Lambda}}$ does not affect the relevance and the ``dominance'' analysis.
The relevance condition requires negative exponent of $T$ in Eq.~(\ref{chiCS2NT}),
\begin{eqnarray}
	&& \gamma > \frac{D + 1}{N} - \frac{D + 3}{2 N^2} \, . \label{2NKFrelevance}
\end{eqnarray}
Using the series representation of $\mathcal{A}_{\mathrm{C/S, 2N}}\left(\gamma\right)$,
\begin{eqnarray}
	&& \hspace{-35pt} \mathcal{A}_{\mathrm{C/S, 2N}}\left(\gamma\right) = \sum\limits_{n = 0}^\infty \left[\frac{\Gamma\left(n + \frac{\displaystyle \alpha_{\mathrm{C/S}}}{\displaystyle 2}\right)}{\Gamma\left(\frac{\displaystyle \alpha_{\mathrm{C/S}}}{\displaystyle 2}\right) n!}\right]^2 \!\!\! \frac{\Gamma\left(1 - a\right)}{\left(2 n + N\right)^{1 - a}} \, , \label{ACS2Nseries} \\ [5pt]
	&& \hspace{-35pt} a = \left(N - \frac{1}{2}\right)\left(D - 1\right) \, ,
\end{eqnarray}
it is easy to show that Eq.~(\ref{2NKFrelevance}) ensures convergence of the series.
Taking into account that the regular part of the composite density wave susceptibility scales as $k_{\mathrm{F}}^{2 N D - D - 1}/v_{\mathrm{F}}$, we find that the non-analytic part is dominant at small temperatures $T \sim \Lambda_{\mathrm{IR}}$ if $\gamma$ satisfies the following condition,
\begin{eqnarray}
	&& \gamma > 2\frac{D + 1}{N} - \frac{D + 3}{N^2} \, . \label{2NKFdominance}
\end{eqnarray}
We also point out that if this ``dominance'' condition is satisfied, then $\gamma > \gamma_{\mathrm{2N}}$, i.e. the $2Nk_{\mathrm{F}}$ composite susceptibility is automatically the most relevant.
For example, if $N = 2$, then Eq.~(\ref{2NKFdominance}) requires $\gamma > (3 D + 1)/4$.

\section{Conclusions}
\label{sec:conclusions}

We applied the dimensional reduction and utilized the multidimensional bosonization scheme for a $D$-dimensional electron system with a spherical Fermi surface and forward-scattering density-density interaction. 
For non-singular finite-range interactions, the fermion Green function retains its Fermi-liquid form, while susceptibilities pick up a relevant power-law enhancement [pairing (density-wave) susceptibilities for attractive (repulsive) interactions].
We show explicitly that these results are equivalent to the one-loop RG treatment, see Ref.~\cite{miserevMicroscopicMechanismPair2024}.
Multidimensional bosonization allows for a straightforward generalization to study composite susceptibilities probing $2 N e$ superconductivity for attractive and $2 N k_{\mathrm{F}}$ density waves for repulsive finite-range interactions.
We find that at large enough coupling constant, such composite susceptibilities outweigh conventional ``quadratic'' Cooper pair and $2 k_{\mathrm{F}}$ density-wave responses, indicating possible quantum phase transitions to exotic $2 N e$ superconductors or composite $2 N k_{\mathrm{F}}$ density waves.

\section{Acknowledgments}
This work was supported by the Georg H. Endress Foundation and the Swiss National Science Foundation (SNSF). 
D.L. acknowledges the Deanship of Research and the Quantum Center at KFUPM for the support received under Grant no. CUP25102 and no. INQC2600, respectively.

\appendix

\section{Dimensional Reduction}
\label{sec:dimred}

First, we recapitulate the dimensional reduction approach for an interacting $D$-dimensional electron gas with arbitrary forward-scattering density-density interaction.
This Appendix aims to include key results of the dimensional reduction approach that are otherwise scattered over Refs.~\cite{miserevDimensionalReductionLuttingerWard2023a,miserevMicroscopicMechanismPair2024,miserev_high-temperature_2025}.

This approach maps Feynman diagrams with dressed interaction lines and dressed fermion propagators (so-called ``skeleton'' diagrams) onto corresponding 1D skeleton diagrams. 
The elementary diagrammatic objects here are fermion lines (chains of dressed fermion propagators separated by bare interaction vertices) and fermion loops (closed fermion lines).

First, let us start with a $D$-dimensional single-particle Matsubara Green function $G(\tau, \bm r)$, where $\tau$ is an imaginary time, and $\bm r$ is a $D$-dimensional coordinate vector,
\begin{eqnarray}
	&& G(\tau, \bm r) = \int \frac{d \bm p}{(2 \pi)^D} \, e^{i \bm p \cdot \bm r} G(\tau, \bm p) \, , \label{Gdef}
\end{eqnarray}
where $G(\tau, \bm p)$ is the $D$-dimensional Fourier transform, $\bm p$ the $D$-dimensional momentum.
We are interested in asymptotic behavior of $G(\tau, \bm r)$ at $r \gg \lambda_{\mathrm{F}}$ and $\tau \gg 1/E_{\mathrm{F}}$, where $\lambda_{\mathrm{F}} = 2 \pi/ k_{\mathrm{F}}$ is the Fermi wavelength, $E_{\mathrm{F}}$ the Fermi energy.
The leading contribution stems from non-analytic or singular behavior of $G(\tau, \bm p)$ near the Fermi surface: analytic part of $G(\tau, \bm p)$ does not contribute in power-law asymptotics of $G(\tau, \bm r)$ at $r \to \infty$, see Paley-Wiener-Schwartz theorems on Fourier-Laplace transforms of analytic functions.
Here, the Fermi surface is a manifold in the momentum space determining points of non-analyticities/singularities of $G(\tau, \bm p)$ at large $\tau \gg 1/E_{\mathrm{F}}$.
The Fermi surface can be also envisioned as a manifold where the fermion occupation number fluctuates the most.
The condition $\tau \gg 1/E_{\mathrm{F}}$ explicitly highlights the low-energy sector. 
In this work, we are not interested in non-analyticities/singularities that are buried deep below or high above the Fermi surface, as such singularities, even when possible, are not relevant to quantum phase transitions.

In order to derive the long-wavelength infrared asymptotics of $G(\tau, \bm r)$, we apply the stationary phase method.
This method has been applied for an arbitrary smooth Fermi surface in Ref.~\cite{miserev_instability_2022}.
Here, we only show the derivation for a spherical Fermi surface with the Fermi momentum $k_{\mathrm{F}}$, in order to demonstrate the key features of this approach.
As the leading contribution in Eq.~(\ref{Gdef}) comes from near the Fermi surface, we approximate the integration region by a thin shell around the Fermi surface, $\bm p = \bm k + q \bm n_{\bm k}$, where $\bm k$ is a point on the Fermi surface that is closest to $\bm p$, $\bm n_{\bm k}$ is the outward normal to the Fermi surface at $\bm k$, $q$ is the distance from $\bm p$ to the Fermi surface,
\begin{eqnarray}
	&& \hspace{-25pt} G(\tau, \bm r) \approx \int\limits_{-\infty}^\infty \frac{d q}{2 \pi} \int\limits_{\mathrm{FS}} \frac{d \bm k}{(2 \pi)^{D - 1}} e^{i (\bm k + q \bm n_{\bm k}) \cdot \bm r} G_{\bm k}(\tau, q) \, , \label{GFS}
\end{eqnarray}
where $G_{\bm k}(\tau, q) = G(\tau, \bm k + q \bm n_{\bm k})$.
For a spherical Fermi surface,  $\bm n_{\bm k} = \bm k/k_{\mathrm{F}}$, $k_{\mathrm{F}} = |\bm k|$, $G_{\bm k}(\tau, q)$ has a non-analyticity/singularity at $q \to 0$ meanwhile its $\bm k$ dependence is assumed to be regular [the same kind of non-analyticity/singularity at any point $\bm k$ on the Fermi surface].
Fast oscillations come from the factor $e^{i \bm k\cdot \bm r}$ in Eq.~(\ref{GFS}), therefore the stationary phase is defined by $d \bm k \cdot \bm r = 0$, where $\bm k$ and $\bm k + d \bm k$ are two infinitesimally close wave-vectors on the Fermi surface, i.e. $d \bm k$ is an element of the tangent space to the Fermi surface at point $\bm k$.
Ergo, the stationary points are all points on the Fermi surface whose tangent spaces are orthogonal to the coordinate vector $\bm r$.
In case of a spherical Fermi surface, there are always two stationary points: $\bm k_\nu = \nu k_{\mathrm{F}} \bm r/r$ with $\bm n_{\bm k_\nu} = \nu \bm r/r$, where $\nu \in \{\pm 1\}$.
Therefore, the leading-order contribution to $G(\tau, \bm r)$ comes from small vicinity of two stationary points $\bm k_\nu$,
\begin{eqnarray}
	&& \hspace{-25pt} G(\tau, \bm r) \approx \sum\limits_{\nu = \pm 1} e^{i \nu k_{\mathrm{F}} r}  \int\limits_{-\infty}^\infty \frac{d q}{2 \pi} \, G_{\bm k_\nu}(\tau, q) e^{i \nu q r} \nonumber \\
	&&  \hspace{7pt} \times \int\frac{d \boldsymbol{\kappa}}{(2 \pi)^{D - 1}} e^{i \boldsymbol{\kappa} \cdot \bm r} \, ,
\end{eqnarray}
where $\boldsymbol{\kappa}  = \bm k - \bm k_{\nu}$ is a parametrization of the Fermi surface near the stationary point $\bm k_{\nu}$, and the integral over $\boldsymbol{\kappa}$ is taken over a small vicinity of $\boldsymbol{\kappa}=0$.
For a spherical Fermi surface, we find that $k_{\mathrm{F}}^2 = (\boldsymbol{\kappa} + \bm k_{\nu})^2$, which yields the following expression,
\begin{eqnarray}
	&& \boldsymbol{\kappa} \cdot \bm r = - \frac{\nu r}{2 k_{\mathrm{F}}} \boldsymbol{\kappa}^2 \, , 
\end{eqnarray}
where we used that $\bm k_\nu = \nu k_{\mathrm{F}} \bm r/r$.
From this, we find that the integral over $\boldsymbol{\kappa}$ is a $(D-1)$-dimensional Gaussian integral,
\begin{eqnarray}
	&& \hspace{-20pt} \int\frac{d \boldsymbol{\kappa}}{(2 \pi)^{D - 1}} e^{i \boldsymbol{\kappa} \cdot \bm r} = \left[\int\limits_{-\infty}^\infty \frac{d \kappa_1}{2 \pi} \exp\left(-i \nu \frac{r \kappa_1^2}{2 k_{\mathrm{F}}}\right)\right]^{D-1} \nonumber \\ [5pt]
	&& \hspace{53pt} = \frac{1}{\left(\lambda_{\mathrm{F}} r\right)^{\frac{D - 1}{2}}} e^{-i \nu \vartheta_{\mathrm{D}}} \, , \\ [5pt]
	&&  \hspace{-20pt} \vartheta_{\mathrm{D}} = \left(D-1\right) \frac{\pi}{4} \, , \label{thetaphase}
\end{eqnarray}
where $\vartheta_{\mathrm{D}}$ plays role of the semiclassical phase.
Here, we expanded $\boldsymbol{\kappa}^2 = \kappa_1^2 + \dots + \kappa_{D - 1}^2$ via $D-1$ principal Euler directions of the spherical Fermi surface.
The Gaussian integral converges quickly on the scale of $|\kappa_j| \sim \sqrt{k_{\mathrm{F}}/r}$, $j \in \{1,\dots, D - 1\}$, corresponding to small angular measure $\delta \theta_j \approx |\kappa_j|/k_{\mathrm{F}} \sim 1/\sqrt{k_{\mathrm{F}} r} \ll 1$ justifying the asymptotic expansion at $k_{\mathrm{F}} r \gg 1$.
This results in the following asymptotic expansion of the fermion Green function,
\begin{eqnarray}
	&& G(\tau, \bm r) \approx \sum\limits_{\nu = \pm 1} \frac{e^{i \nu \left(k_{\mathrm{F}} r - \vartheta_{\mathrm{D}}\right)}}{(\lambda_{\mathrm{F}} r)^{\frac{D - 1}{2}}} g_\nu (\tau, r) \, , \label{Gnu} \\
	&& g_\nu (\tau, r) = \int\limits_{-\infty}^\infty \frac{d q}{2 \pi} \, G_{\bm k_\nu}(\tau, q) e^{i \nu q r} \, . \label{gnu}
\end{eqnarray}
Here, $g_\nu (\tau, r)$ is the effective 1D Green function [we omit $\bm k_{\nu}$ index for brevity].
The index $\nu$ plays role of the chirality index.
We point out that the integral over $q$ in Eq.~(\ref{gnu}) is convergent at $q \sim 1/r \ll k_{\mathrm{F}}$ which justifies expanding the integration limits to infinity.
Equation~(\ref{Gnu}) explicitly separates fast oscillatory exponential, $e^{\pm i k_{\mathrm{F}} r}$, from slow 1D Green function $g_\nu(\tau, r)$.
This scale separation is in the heart of the dimensional reduction.

\begin{figure}[t]
	\centering
	\includegraphics[width=0.99\columnwidth]{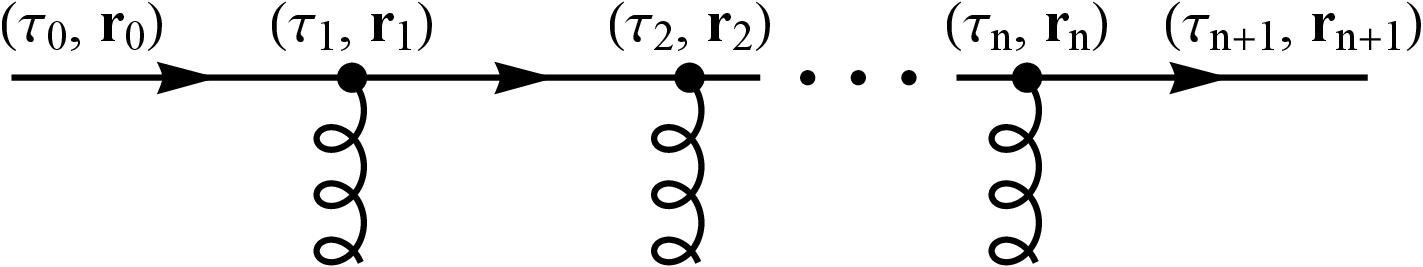} 
	\caption{A fermion line with $n$ internal vertices and two fixed points $(\tau_0, \bm r_0)$ and $(\tau_{n+1}, \bm r_{n+1})$ that we analyze in Eq.~(\ref{Farb}), other possible fermion lines or loops are not shown. Black points represent interaction vertices, directed lines are dressed fermion propagators, wavy lines correspond to dressed $D$-dimensional interaction. Some interaction lines can connect back to the same fermion line or some other fermion line or loop not shown on this figure, the final result given in Eq.~(\ref{Fdimred}) is independent of such details.}
	\label{fig:diag}
\end{figure}

Next, we consider a Feynman diagram containing a fermion line, see Fig.~\ref{fig:diag},
\begin{eqnarray}
	&& \hspace{-25pt} F = \int \left(\dots\right) \int \prod\limits_{j = 1}^n \left[ d\bm r_j d\tau_j \, U_{\mathrm{D}}(\tau_j - T_j, \bm r_j - \bm R_j)\right] \, \nonumber \\
	&& \hspace{33pt} \times \prod\limits_{j = 0}^n \left[ G(\tau_{j+1} - \tau_{j}, \bm r_{j+1} - \bm r_{j}) \right]  \, , \label{Farb}
\end{eqnarray}
where points $(\tau_0, \bm r_0)$, $(\tau_{n+1}, \bm r_{n+1})$ are external points on the fermion line, $(\dots)$ is a part of the diagram unrelated to chosen fermion line, $G(\tau, \bm r)$ is a fermion Green function, see Eq.~(\ref{Gnu}), $U_{\mathrm{D}}(\tau, \bm r)$ is dressed forward-scattering interaction, i.e. we assume that $U_{\mathrm{D}}(\tau, \bm r)$ is a slow function at $r \gg 1/k_{\mathrm{F}}$, $\tau \gg 1/E_{\mathrm{F}}$. 
Some points $(T_j, \bm R_j)$ might coincide with $(\tau_{j'}, \bm r_{j'})$ if corresponding interaction line starts and ends on the chosen fermion line.
Fast oscillatory factors originate from the product of Green functions in Eq.~(\ref{Farb}), therefore slow interaction functions do not affect stationary points for the integrals over $\bm r_1, \dots, \bm r_n$,
\begin{eqnarray}
	&& \hspace{-28pt} \prod\limits_{j = 0}^n \left[ G(\tau_{j+1} - \tau_{j}, \bm r_{j+1} - \bm r_{j}) \right] \approx \sum\limits_{\nu_0,\dots,\nu_n} e^{-i \vartheta_{\mathrm{D}} \sum\limits_{j = 0}^n \nu_j}  \nonumber \\
	&& \hspace{-28pt} \times \prod\limits_{j = 0}^n \left[ \frac{g_{\nu_j}(\tau_{j+1} - \tau_{j}, |\bm r_{j+1} - \bm r_{j}|)}{\left(\lambda_{\mathrm{F}} |\bm r_{j+1} - \bm r_{j}|\right)^{\frac{D - 1}{2}}} \right] e^{i \Theta} \, , \label{prodG} \\
	&& \hspace{-28pt} \Theta \equiv k_{\mathrm{F}} \sum\limits_{j = 0}^n \nu_j |\bm r_{j+1} - \bm r_{j}| \, .
\end{eqnarray}
At this point, it is convenient to introduce shifted coordinates,
\begin{eqnarray}
	&& \hspace{-20pt} \bm r_{j\,0} \equiv \bm r_j - \bm r_0 \, , \hspace{5pt} \bm n_{j\,0} \equiv \frac{\bm r_{j\,0}}{r_{j\,0}} \, , \hspace{5pt} r_{j\,0} = \left|\bm r_{j\,0}\right| \, , \label{rj0} 
\end{eqnarray}
where $j \in \{1,\dots,n+1\}$ and $\bm r_{n+1 \, 0} = \bm r_{n+1} - \bm r_0$ is fixed.
The fast oscillatory phase factor $\Theta$ in Eq.~(\ref{prodG}) can thus be written as follows,
\begin{eqnarray}
	&& \frac{\Theta}{k_{\mathrm{F}}} = \nu_0 r_{1\, 0} + \sum\limits_{j = 1}^n \nu_j \left|\bm r_{j\, 0} - \bm r_{j+1 \, 0}\right| \, . \label{Theta}
\end{eqnarray}
Now, we are ready to integrate over the angular variables $\bm n_{j\, 0}$, $j \in \{1,\dots,n\}$ using the stationary phase method.
The stationary points $\bm n_{j\, 0}^*$ are determined as local extrema of the oscillatory phase $\Theta$,
\begin{eqnarray}
	&& \left. \frac{\partial \Theta}{\partial \bm n_{j\,0}}\right|_{\bm n_{j\, 0}^*} = 0 \, .
\end{eqnarray}
This condition is equivalent to that all unit vectors $\bm n_{j\, 0}^*$, $j \in \{1,\dots, n\}$, are collinear to fixed unit vector $\bm n_{n+1 \, 0}$,
\begin{eqnarray}
	&& \bm n_{j \, 0}^* = \sigma_j \bm n_{n+1 \, 0} \, , \hspace{5pt} \sigma_j = \pm 1 \, . \label{ncollin}
\end{eqnarray}
Different sets of stationary points correspond to different sign choices $\sigma_j \in \{\pm 1\}$ in Eq.~(\ref{ncollin}).
For a given choice of $\sigma_j$, the stationary-point value of the phase $\Theta$ is generally dependent on $r_{j \, 0}$,
\begin{eqnarray}
	&& \frac{\Theta^*}{k_{\mathrm{F}}} = \nu_0 r_{1 \, 0} + \nu_n \left|\sigma_n r_{n \, 0} - r_{n+1 \, 0}\right| \nonumber \\
	&& + \sum\limits_{j = 1}^{n-1} \nu_j \left|\sigma_j r_{j \, 0} - \sigma_{j+1} r_{j+1 \, 0}\right| \, , \label{Thetastar}
\end{eqnarray}
where we explicitly take $j = n$ term out of the sum.
We point out that we sum not only over the stationary-point configurations $\{\sigma_j\}$ but also over all indices $\{\nu_0,\dots,\nu_n\}$, see Eq.~(\ref{prodG}).
Any residual linear dependence of $\Theta^*$ on external coordinates $r_{j \, 0} = |\bm r_{j \, 0}|$ results in fast oscillatory phase factor in Eq.~(\ref{prodG}) making such contributions irrelevant.
Therefore, the largest contribution to the diagram $F$, see Eq.~(\ref{Farb}), comes from configurations $\{\sigma_j\}$, $\{\nu_j\}$ corresponding to $\Theta^*$ that is independent of all $r_{j \, 0}$ with $j \in \{1,\dots,n\}$, which is equivalent to the following sign choices,
\begin{eqnarray}
	&& \nu_j = \sigma_1 \nu_0 \, \mathrm{sgn}\left(\sigma_{j+1} r_{j+1 \, 0} - \sigma_j r_{j \, 0}\right) \, , \label{nu} \\
	&& \Theta^* = \sigma_1 \nu_0 k_{\mathrm{F}} r_{n+1 \, 0} \, , \label{Thetamain}
\end{eqnarray}
where $\mathrm{sgn}(x)$ returns the sign of $x$, $r_{n+1 \, 0}$ is fixed, $\sigma_{n+1} = 1$ [see Eq.~(\ref{ncollin})], and $\nu_0$ is the only chirality index remaining free.
It is convenient to introduce new notations,
\begin{eqnarray}
	&& x_j = \sigma_j r_{j\,0} \, , \hspace{5pt} \nu = \sigma_1 \nu_0 \, ,
\end{eqnarray}
where variables $x_j$ play role of one-dimensional coordinates.
The leading contribution to the diagram $F$ is the sum over the stationary-point contributions,
\begin{widetext}
	\begin{eqnarray}
		&& \hspace{-10pt} F \approx \int \left(\dots\right) \int \prod\limits_{j = 1}^n \left[d\tau_j \, d r_{j \, 0}^{\vphantom{D}} \, r_{j \, 0}^{D - 1}\right] \sum\limits_{\nu_0} \sum\limits_{\{\sigma_j\}} e^{i \nu k_{\mathrm{F}} r_{n+1 \, 0} - i \vartheta_{\mathrm{D}} \sum\limits_{j = 0}^n \nu_j} 
		\prod\limits_{j=1}^n \left[ U_{\mathrm{D}}\left(\tau_j-T_j, r_{j \, 0} \bm n_{j \, 0}^* - \bm R_{j \, 0}\right) \right] \nonumber \\
		&& \times \prod\limits_{j = 0}^n \left[\frac{g_{\nu}\left(\tau_{j+1} - \tau_{j}, x_{j+1} - x_{j}\right)}{\left(\lambda_{\mathrm{F}} \left|x_{j+1} - x_{j}\right|\right)^{\frac{D - 1}{2}}}\right] \int \prod\limits_{j = 1}^n \left[d \bm n_{j \, 0}\right] \, e^{i \left(\Theta - \Theta^*\right)} \, , \label{Fapprox}
	\end{eqnarray}
\end{widetext}
where $x_0 = 0$, $\bm n_{j \, 0}^*$, $\nu_j$ are given by Eqs.~(\ref{ncollin}), and (\ref{nu}), respectively, $\bm R_{j \, 0} = \bm R_j - \bm r_0$, $\Theta^*$ is the stationary-point value of $\Theta$, see Eqs.~(\ref{Theta}), (\ref{Thetamain}). 
Here, we used $g_{\nu_1\nu_2}(\tau, r) = g_{\nu_1}(\tau, \nu_2 r)$ which follows from Eq.~(\ref{gnu}).
The phase difference near the stationary point is approximately given by the following quadratic form,
\begin{eqnarray}
	&& \hspace{-35pt} \frac{\Theta - \Theta^*}{k_{\mathrm{F}}}  \approx \sum\limits_{j = 1}^n \nu_j \frac{x_j x_{j+1}\left(\sigma_j \delta \bm n_{j \, 0} - \sigma_{j + 1} \delta \bm n_{j + 1 \, 0}\right)^2}{2 \left|x_j - x_{j+1}\right|}  \, , \label{dTheta}
\end{eqnarray}
where $\delta \bm n_{j \, 0} = \bm n_{j \, 0} - \bm n_{j \, 0}^*$.
It is now convenient to introduce the angle between $\sigma_j \bm n_{j \, 0}$ and $\sigma_{j+1}\bm n_{j+1 \, 0}$,
\begin{eqnarray}
	&& \cos \theta_j \equiv \sigma_j \sigma_{j+1} \bm n_{j \, 0} \cdot \bm n_{j + 1 \, 0} \nonumber \\
	&& = 1 - \frac{\left(\sigma_j \delta \bm n_{j \, 0} - \sigma_{j + 1} \delta \bm n_{j + 1 \, 0}\right)^2}{2} \, .
\end{eqnarray}
Notice that $\theta_j = 0$ right at the stationary point due to Eq.~(\ref{ncollin}). Therefore, $\theta_j \ll 1$ in a small vicinity of the stationary point, which gives the approximate relation,
\begin{eqnarray}
	&& \theta_j^2 \approx \left(\sigma_j \delta \bm n_{j \, 0} - \sigma_{j + 1} \delta \bm n_{j + 1 \, 0}\right)^2 \, .
\end{eqnarray}
This allows us to rewrite Eq.~(\ref{dTheta}) as follows,
\begin{eqnarray}
	&& \Theta - \Theta^* \approx k_{\mathrm{F}} \sum\limits_{j = 1}^n \frac{\nu_j x_j x_{j+1} \, \theta_j^2}{2 \left|x_j - x_{j+1}\right|}  \, . \label{dTheta2}
\end{eqnarray}
Using Eq.~(\ref{dTheta2}) and $d\bm n \approx S_{\mathrm{D - 2}} \theta^{D-2} \, d\theta$ for the angular measure at $\theta \ll 1$, where $S_{\mathrm{d}}$ is the surface area of a unit $d$-dimensional sphere, we can now calculate the angular integrals in Eq.~(\ref{Fapprox}),
\begin{eqnarray}
	&& \hspace{-20pt} \int \prod\limits_{j = 1}^n \left[d \bm n_{j0}\right] \, e^{i \left(\Theta - \Theta^*\right)} \nonumber \\
	&& \hspace{-20pt} \approx \prod\limits_{j = 1}^n \left[S_{\mathrm{D - 2}} \int\limits_0^\infty d\theta_j \, \theta_j^{D-2} \exp \left(\frac{i \nu_j k_{\mathrm{F}} x_j x_{j + 1} \theta_j^2}{2 |x_j - x_{j+1}|}\right)\right] \nonumber \\
	&& \hspace{-20pt} = \prod\limits_{j = 1}^n \left[e^{i \vartheta_{\mathrm{D}} \nu_j \sigma_j \sigma_{j+1}} \left(\frac{\lambda_{\mathrm{F}} |x_j - x_{j+1}|}{r_{j\,0} r_{j+1 \, 0}}\right)^{\frac{D-1}{2}} \right] \, .
\end{eqnarray}
Here, we used quick convergence of integrals over $\theta_j$ and extended the integration interval to infinity.
Substituting this back into Eq.~(\ref{Fapprox}), we find,
\begin{eqnarray}
	&& F \approx \int \left(\cdots\right) \sum\limits_{\nu = \pm 1} \frac{e^{i \nu k_{\mathrm{F}} r_{n+1\, 0}}}{\left(\lambda_{\mathrm{F}} r_{n+1\, 0}\right)^{\frac{D-1}{2}}} \sum\limits_{\{\sigma_j\}} e^{-i \vartheta_{\mathrm{D}} \alpha} \nonumber \\
	&& \times \int\limits_0^{\infty} \prod\limits_{j = 1}^{n} \left[d\tau_j \, dr_{j \, 0} \, U_{\mathrm{D}}\left(\tau_j -T_j, x_j \bm n_{n+1\, 0}- \bm R_{j\, 0}\right)\right] \nonumber \\
	&& \times \prod\limits_{j = 0}^{n} \left[g_\nu \left(\tau_{j+1} - \tau_{j}, x_{j+1} - x_{j}\right)\right] \, , \label{Fapprox2} \\
	&& \alpha \equiv \nu_0 + \sum\limits_{j = 1}^n \nu_j \left(1 - \sigma_j \sigma_{j + 1}\right) \, , \label{alpha}
\end{eqnarray}
where $x_0 = 0$, and $r_{j\, 0} \in (0, +\infty)$ which is indicated by integration limits. 
In order to simplify expression for $\alpha$, we notice that $\nu_j$ contributes to the sum only if $\sigma_j \sigma_{j+1} = -1$. From Eq.~(\ref{nu}), we find that $\nu_j = \nu \sigma_{j+1}$ at this condition. This yields the telescoping sum in Eq.~(\ref{alpha}) resulting in $\alpha = \nu$, which is independent of $\sigma_j$.
As the integrand in Eq.~(\ref{Fapprox2}) depends on $\sigma_j$ only via $x_j$, the sum over $\{\sigma_j\}$ just extends the integration over each 1D coordinate $x_j$ over the whole real line, i.e. $x_j \in (-\infty, \infty)$.
Therefore, the diagram $F$ takes the following dimension-reduced form,
\begin{eqnarray}
	&& F \approx \int \left(\cdots\right) \sum\limits_{\nu = \pm 1} \frac{e^{i \nu \left( k_{\mathrm{F}} r_{n+1\, 0} - \vartheta_{\mathrm{D}} \right)}}{\left(\lambda_{\mathrm{F}} r_{n+1\, 0}\right)^{\frac{D-1}{2}}} \nonumber \\
	&& \times \int\limits_{-\infty}^{\infty} \prod\limits_{j = 1}^{n} \left[dx_{j} \, d\tau_j \, U_{\mathrm{D}}\left(\tau_j -T_j, x_j \bm n_{n+1\, 0}- \bm R_{j\, 0}\right)\right] \nonumber \\
	&& \times \prod\limits_{j = 0}^{n} \left[g_\nu \left(\tau_{j+1} - \tau_{j}, x_{j+1} - x_{j}\right)\right] \, , \label{Fdimred} 
\end{eqnarray}
where $x_0 = 0$, $x_{n+1} = r_{n+1 \, 0} = |\bm r_{n+1} - \bm r_0|$, $\bm n_{n+1 \, 0} = \bm r_{n+1 \, 0}/r_{n+1 \, 0}$.
Equation~(\ref{Fdimred}) represents the map from a $D$-dimensional ``skeleton'' diagram containing fermion lines onto the 1D diagram of the same structure.
The only important difference from 1D case is $1/r^{\frac{D-1}{2}}$ factor per fermion line. This power-law factor originating from the Fermi-surface curvature is crucial for studying quantum phase transitions in $D$-dimensional interacting fermion systems.

We showed that fermion lines are mapped onto corresponding 1D lines by the rule in Eq.~(\ref{Fdimred}).
This rule also applies to the fermion loops representing closed fermion lines. 
Therefore, the dimensional reduction allows us to use the fermion loop cancellation theorem that is exact for 1D fermion systems with linear dispersion and forward-scattering interaction \cite{dzyaloshinskiilarkin}. 
This theorem states that symmetrized 1D fermion loops with three or more forward-scattering interaction vertices are canceled exactly.
More general study of $D$-dimensional symmetrized fermion loops shows that they remain finite in the static limit (frequencies to zero) and vanish in the dynamic limit (momentum transfers to zero) \cite{neumayr_fermion_1998}. 
Here, what is important is that symmetrized fermion loops are not singular in any spatial dimensions, and therefore diagrams containing fermion loops are irrelevant compared to the diagrams without fermion loops.
The only fermion loop that still provides a sizable non-analyticity is the particle-hole bubble dressing the interaction. 
In particular, dressed $D$-dimensional interaction at small frequency and momentum is well described by RPA.

\section{RPA Interaction}
\label{sec:RPA}

In this section, we recall the $D$-dimensional RPA screening focusing on the static and dynamic limits. This Appendix contains well-known results and is included to make the paper easier to read.

The $D$-dimensional polarization operator $\Pi_{\mathrm{D}}$ at small frequency and momentum can be approximated by the particle-hole bubble $\Pi_{\mathrm{D}}^{(0)}$ that can be best represented in the space-time coordinates as follows,
\begin{eqnarray}
	&& \hspace{-22pt} \Pi_{\mathrm{D}}(\tau, r) \approx \Pi_{\mathrm{D}}^{(0)}(\tau, r) = N G^{(0)}(\tau, r) G^{(0)}(-\tau, r) \, , \label{PiD}
\end{eqnarray}
where $N$ is the number of fermion flavors, $G^{(0)}(\tau, r)$ is the free electron Green function with the long-distance asymptotics given by Eq.~(\ref{Gnu}).
Substituting the asymptotics of the Green function back into Eq.~(\ref{PiD}), we find,
\begin{eqnarray}
	&& \hspace{-25pt} \Pi_{\mathrm{D}}(\tau, r) \approx \frac{N}{(\lambda_{\mathrm{F}} r)^{D - 1}} \nonumber \\
	&& \hspace{-25pt} \times \sum\limits_{\nu, \nu' = \pm 1} e^{i (\nu + \nu')(k_{\mathrm{F}} r - \vartheta_{\mathrm{D}})} g^{(0)}_{\nu}(\tau, r) g^{(0)}_{-\nu'}(-\tau, -r) \, , \label{PiDasympt}
\end{eqnarray}
where we used that $g^{(0)}_{\nu'}(-\tau, x) = g^{(0)}_{-\nu'}(-\tau, -x)$.
We are interested in the small-momentum behavior of the polarization operator given by $\nu' = -\nu$ contribution in Eq.~(\ref{PiDasympt}).
The contribution at $\nu' = \nu$ corresponds to the $2 k_{\mathrm{F}}$ Kohn anomaly of the polarization operator, and is not related to the forward-scattering physics we are studying here.
Therefore, the contribution of our interest can be written as follows, 
\begin{eqnarray}
	&& \Pi_{\mathrm{D}}(\tau, r) \approx -\frac{N}{(\lambda_{\mathrm{F}} r)^{D - 1}} \sum\limits_{\nu = \pm 1} \left[g^{(0)}_\nu(\tau, r)\right]^2 \, ,
\end{eqnarray}
where we used that $g^{(0)}_\nu(-\tau, -x) = - g^{(0)}_\nu(\tau, x)$.
The free fermion Green function is given by the following expression,
\begin{eqnarray}
	&& \hspace{-25pt} G^{(0)}(i \omega_n, q) \approx \frac{1}{i \omega_n - q v_{\mathrm{F}}} \, , \label{G0wq} \\
	&& \hspace{-25pt} g_\nu^{(0)} (\tau, x) = T \sum\limits_{\omega_n} \int\limits_{-\infty}^\infty \frac{dq}{2 \pi} \, e^{i q \nu x - i \omega_n \tau} G^{(0)}(i \omega_n, q) \nonumber \\
	&& \hspace{17pt} = - \frac{\displaystyle T}{2 v_{\mathrm{F}} \displaystyle \sin\left[\pi T \left(\tau - i \nu \frac{x}{v_{\mathrm{F}}}\right)\right]} \, , \label{g0taux}
\end{eqnarray}
where $q = p - k_{\mathrm{F}} \ll k_{\mathrm{F}}$ is the distance to the Fermi surface, $v_{\mathrm{F}}$ the Fermi velocity, $\omega_n$ the fermionic Matsubara frequency, $T$ the temperature.

Now, we are ready to take the Fourier transform at small momentum $p$ and small Matsubara frequency $\omega_n$,
\begin{eqnarray}
	&& \hspace{-20pt} \Pi_{\mathrm{D}}(i \omega_n, p) \nonumber \\
	&& \hspace{-20pt} = \int\limits_{0}^\beta d\tau \int\limits_0^\infty dr\, r^{D-1} e^{i \omega_n \tau} \Pi_{\mathrm{D}}(\tau, r) \int d\bm n \, e^{- i \bm p \cdot \bm r} \, ,
\end{eqnarray}
where $\beta = 1/T$.
The angular integral can be expressed in terms of the Bessel function $J_\mu(z)$ of the first kind,
\begin{eqnarray}
	&& \hspace{-20pt} \int d\bm n \, e^{-i \bm p \cdot \bm r} = 2 \pi \left(\frac{2 \pi}{p r}\right)^{\frac{D}{2}-1} J_{\frac{D}{2}-1}(pr) \, , \label{angleint} \\
	&&  \hspace{-20pt} J_\mu (z) = \frac{z^\mu}{2^\mu \sqrt{\pi} \Gamma\left(\mu + \frac{1}{2}\right)} \int\limits_{-1}^1 ds \, (1 - s^2)^{\mu - \frac{1}{2}} e^{i s z} \, , \label{Bessel}
\end{eqnarray}
where $\Gamma(z)$ is the gamma function.
Substituting the Bessel function from Eq.~(\ref{Bessel}), we find,
\begin{eqnarray}
	&& \hspace{-30pt} \Pi_{\mathrm{D}}(i \omega_n, p) = - N_{\mathrm{F}} v_{\mathrm{F}} \frac{\sqrt{\pi} \Gamma\left(\frac{D}{2}\right)}{\Gamma\left(\frac{D-1}{2}\right)} \int\limits_{-1}^1 ds \, (1-s^2)^{\frac{D-3}{2}} \nonumber \\
	&& \hspace{-5pt} \times \int\limits_{0}^\beta d\tau\int\limits_{-\infty}^\infty dr \, e^{i \omega_n \tau + i s p r} \sum\limits_{\nu = \pm 1} \left[g_\nu^{(0)}(\tau, r)\right]^2 \, , \label{PiDs}
\end{eqnarray}
where $N_{\mathrm{F}}$ is the total density of states at the Fermi level,
\begin{eqnarray}
	&& N_{\mathrm{F}} = \frac{N}{\lambda_{\mathrm{F}}^{D-1} v_{\mathrm{F}}} \frac{\pi^{\frac{D-2}{2}}}{\Gamma\left(\frac{D}{2}\right)} \, . \label{NF}
\end{eqnarray}
We also extended integration over $r$ in Eq.~(\ref{PiDs}) to the real line $(-\infty, \infty)$.
The integral over $\tau$ and $r$ corresponds to the 1D particle-hole bubble,
\begin{eqnarray}
	&& \hspace{-30pt} \int\limits_{0}^\beta d\tau\int\limits_{-\infty}^\infty dr \, e^{i \omega_n \tau - i s p r} \sum\limits_{\nu = \pm 1} \left[g_\nu^{(0)}(\tau, r)\right]^2  \nonumber \\
	&& = \frac{1}{\pi v_{\mathrm{F}}} \frac{s^2}{s^2 + \left(\omega_n/pv_{\mathrm{F}}\right)^2} \, .
\end{eqnarray}
Substituting this back into Eq.~(\ref{PiDs}), and introducing a new variable $u = s^2$, we find the $D$-dimensional polarization operator,
\begin{eqnarray}
	&& \hspace{-30pt} \Pi_{\mathrm{D}}(i \omega_n, p) =- \frac{N_{\mathrm{F}} \Gamma\left(\frac{D}{2}\right)}{\sqrt{\pi}\Gamma\left(\frac{D-1}{2}\right)} \int\limits_{0}^1 du \, \frac{u^{\frac{1}{2}} (1-u)^{\frac{D-3}{2}}}{u + \zeta^2} \, , \label{PiDexplicit} \\
	&& \hspace{-30pt} \zeta = \frac{\omega_n}{v_{\mathrm{F}} p} \, . \label{zeta}
\end{eqnarray}
Using the Euler integral representation of the hypergeometric function $_2F_1(a,b;c;z)$, we can represent the polarization operator as follows,
\begin{eqnarray}
	&& \hspace{-35pt} \Pi_{\mathrm{D}}(i \omega_n, p) \nonumber \\
	&& \hspace{-35pt}  = -\frac{N_{\mathrm{F}}}{D} \left(\frac{v_{\mathrm{F}} p}{\omega_n}\right)^2 {}_2F_1\left(1,\frac{3}{2};\frac{D}{2}+1;-\left(\frac{v_{\mathrm{F}} p}{\omega_n}\right)^2\right) \, . \label{PiDhyper}
\end{eqnarray}
We point out that this expression is valid at finite temperature $T > 0$.

If $D \in \{1,2,3\}$, the hypergeometric function in Eq.~(\ref{PiDhyper}) can be expressed in terms of elementary functions,
\begin{eqnarray}
	&& \hspace{-25pt} \Pi_{\mathrm{1D}}(i \omega_n, p) = - \frac{N_{\mathrm{F}}}{1 + \zeta^2} \, , \\
	&& \hspace{-25pt} \Pi_{\mathrm{2D}} (i \omega_n, p) = -N_{\mathrm{F}}\left(1 - \frac{\left|\zeta\right|}{\sqrt{1 + \zeta^2}}\right) \, , \\
	&& \hspace{-25pt} \Pi_{\mathrm{3D}} (i \omega_n, p) = -N_{\mathrm{F}} \left(1 - \zeta \arctan\left(\frac{1}{\zeta}\right)\right) \, ,
\end{eqnarray}
where $\zeta$ is given by Eq.~(\ref{zeta}).
These expressions can also be used on real frequencies $\omega$ with the help of analytic continuation $i \omega_n \to \omega + i0^+$.
We point out that in all these cases the polarization operator is singular at real frequencies $\omega = \pm v_{\mathrm{F}} p$, i.e. when $\zeta^2 = -1$.

The $D$-dimensional RPA interaction $U_{\mathrm{D}}(i \omega_n, p)$ is dressed by the polarization operator $\Pi_{\mathrm{D}}(i \omega_n, p)$,
\begin{eqnarray}
	&& \hspace{-20pt} U_{\mathrm{D}} (i \omega_n, p) = \left[U_0^{-1}(i \omega_n, p) - \Pi_{\mathrm{D}}(i \omega_n, p)\right]^{-1} \, , \label{URPA}
\end{eqnarray}
where $U_0(i \omega_n, p)$ is the bare $D$-dimensional interaction.

\section{Luttinger liquids in higher dimensions}
\label{sec:LL}

In this Appendix, we reverse engineer dressed $D$-dimensional interactions resulting in the Luttinger liquid form of the electron spectral function.

First, we start from 1D Luttinger liquids produced by the bare interaction $V_0(i \omega_n, q) = \mathrm{const}$ at $|q v_{\mathrm{F}}| \ll \Lambda$,
where $\Lambda \ll E_{\mathrm{F}}$ is the UV cutoff.
Dressed 1D Luttinger-liquid interaction follows from Eqs.~(\ref{Vgen}), (\ref{dPi1D}) [see Ref.~\cite{dzyaloshinskiilarkin}],
\begin{eqnarray}
	&& V_{\mathrm{LL}}(i \omega_n, q) = V_0 \frac{\omega_n^2 + (q v_{\mathrm{F}})^2}{\omega_n^2 + (q v)^2} \, , \label{U1D} \\
	&& K_{\mathrm{LL}} = \frac{v_{\mathrm{F}}}{v} = \left(1 + \frac{N V_0}{\pi v_{\mathrm{F}}}\right)^{-\frac{1}{2}} \, , \label{KLL}
\end{eqnarray}
where $v$ is the renormalized velocity of the zero sound, $K_{\mathrm{LL}}$ is the Luttinger liquid parameter.
We point out that Eq.~(\ref{GMetzner}) is equivalent to its regularized version, Eq.~(\ref{GMetznerreg}), as evidently $V_{\mathrm{LL}}(\pm q v_{\mathrm{F}}, q) = 0$.
Using Eq.~(\ref{GMetzner}), we then find $\mathcal{L}_+(\xi)$,
\begin{eqnarray}
	&& \hspace{-25pt} \mathcal{L}_+(\xi) = \frac{1}{N} \ln\left[\frac{\sinh\left(\pi T \left(\frac{x}{v_{\mathrm{F}}} + i \tau\right)\right)}{\sinh\left(\pi T \left(\frac{x}{v} + i \tau\right)\right)}\right] \nonumber \\
	&& \hspace{-25pt} - 2 \alpha_{\mathrm{LL}} \ln \left[\frac{\Lambda}{\pi T} \left|\sinh\left(\pi T \left(\frac{x}{v} + i \tau\right)\right)\right|\right] \, , \label{Llut} \\
	&& \alpha_{\mathrm{LL}} = \frac{\left(K_{\mathrm{LL}} - 1\right)^2}{4 N K_{\mathrm{LL}}} \, , \label{alphaLL}
\end{eqnarray}
where $\Lambda \sim v_{\mathrm{F}}/R_{\mathrm{s}}$ is the ultraviolet cut-off originating from the constant part of $\mathcal{L}_+(\xi)$ whose convergence is guaranteed by the falloff of $V_0(i \omega_n, q)$ at $q \sim 1/R_{\mathrm{s}}$.
Before performing the analytic continuation, use $|\sinh(z)|^2 = \sinh(z) \sinh(z^*)$, where $z^*$ is the complex conjugate of $z$.
Indeed, we demonstrated that Eq.~(\ref{GMetzner}) and its equivalent representation given by Eqs.~(\ref{gnuexact}), (\ref{Lnu}) reproduce the Luttinger liquid Green function for the finite-range bare interaction $V_0(i \omega_n, q) \approx \mathrm{const.}$ at $q \ll 1/R_{\mathrm{s}}$.

Using Eq.~(\ref{GMetzner}), we find the following relation between $\mathcal{L}_+(\xi)$ and $V(\tau, x)$,
\begin{eqnarray}
	&& V\left(\tau, x\right) = -4 \frac{\partial^2 \mathcal{L}_+(\tau, x)}{\partial \overline{z}^2} \, , \label{VfromL}
\end{eqnarray}
where $\overline{z} = \tau + i x/v_{\mathrm{F}}$, the derivative is taken at constant $z = \tau - i x/v_{\mathrm{F}}$.
The contact $\propto \delta(\tau) \delta(x)$ term in Eq.~(\ref{VfromL}) is omitted: it corresponds to the constant part of $V_{\mathrm{LL}}(i \omega_n, q)$ which can be safely subtracted according to Sec.~\ref{sec:GF}.
We point out that Eq.~(\ref{VfromL}) can be used to engineer interacting fermion systems with desired $\mathcal{L}_+(\tau, x)$ which defines the single-particle spectral function. 
For example, using the Luttinger-liquid form of $\mathcal{L}_+(\tau, x)$ given by Eq.~(\ref{Llut}), we find the long-range part of $V_{\mathrm{LL}}(\tau, x)$,
\begin{eqnarray}
	&& \hspace{-30pt} V_{\mathrm{LL}} (\tau, x) = - \frac{\left(K_{\mathrm{LL}}^2 - 1\right)^2}{4 N K_{\mathrm{LL}}} \nonumber \\
	&& \hspace{11pt} \times \left[\left(\frac{\pi T}{\sin \left(\pi T z_{\mathrm{v}}\right)}\right)^2  + \left(\frac{\pi T}{\sin \left(\pi T \overline{z}_{\mathrm{v}}\right)}\right)^2\right] \, , \label{VLLtaux}
\end{eqnarray}
where $z_{\mathrm{v}} = \tau - i x/v$, $\overline{z}_{\mathrm{v}} = \tau + i x/v$, $v = v_{\mathrm{F}}/K_{\mathrm{LL}}$.
Using Eq.~(\ref{UV}) and taking the finite-temperature Fourier transform, we find the $D$-dimensional interaction $U_{\mathrm{LL}}(i \omega_n, r)$ that provides the Luttinger-liquid spectral function in $D$ dimensions,
\begin{eqnarray}
	&& \hspace{-20pt} U_{\mathrm{LL}}(i \omega_n, r) = \frac{\pi \left(K_{\mathrm{LL}}^2 - 1\right)^2}{2 N K_{\mathrm{LL}}} \left|\omega_n\right| e^{-|\omega_n|r/v} \, .
\end{eqnarray}
The $D$-dimensional Fourier transform yields the following form of the $D$-dimensional ``Luttinger liquid'' interaction,
\begin{eqnarray}
	&& \hspace{-20pt} U_{\mathrm{LL}}(i \omega_n, p) = \frac{C_{\mathrm{LL}} v^D \omega_n^2}{\left[\omega_n^2 + (p v)^2\right]^{\frac{D + 1}{2}}} \, , \label{ULL} \\
	&& \hspace{-20pt} C_{\mathrm{LL}} = \frac{\left(K_{\mathrm{LL}}^2 - 1\right)^2}{4 N K_{\mathrm{LL}}} \left(4 \pi\right)^{\frac{D + 1}{2}} \Gamma\left(\frac{D +1}{2}\right) \, ,
\end{eqnarray}
where $p = |\bm p|$, $\bm p$ is the $D$-dimensional momentum transfer.
We used the following identity to derive Eq.~(\ref{ULL}),
\begin{eqnarray}
	&& \hspace{-20pt} \int\limits_0^\infty du \, u^{\frac{D}{2}} e^{-\lambda u} J_{\frac{D}{2} - 1} (u) = \frac{2^{\frac{D}{2}}}{\sqrt{\pi}} \frac{\Gamma \left(\frac{D + 1}{2}\right) \lambda}{\left[\lambda^2 + 1\right]^{\frac{D + 1}{2}}} \, ,
\end{eqnarray}
where $\mathrm{Re}(\lambda) > 0$, $J_\mu(u)$ is the Bessel function.
In case if $D = 1$, Eq.~(\ref{ULL}) coincides with Eq.~(\ref{U1D}) up to a constant corresponding to the contact part.
One can show that the following deformations of the dressed $D$-dimensional interaction $U_{\mathrm{LL}}(i \omega_n, p)$ also result in logarithmic singularities in $\mathcal{L}_+(\tau, x)$, and therefore can also be identified as Luttinger liquids,
\begin{eqnarray}
	&& \tilde{U}_{\mathrm{LL}}(i \omega_n, p) \propto \frac{p^2}{\left[\omega_n^2 + (p v)^2\right]^{\frac{D + 1}{2}}} \, , \label{Utilde} \\
	&& \tilde{\tilde{U}}_{\mathrm{LL}}(i \omega_n, p) \propto \frac{1}{\left[\omega_n^2 + (p v)^2\right]^{\frac{D - 1}{2}}} \, . \label{U2tilde}
\end{eqnarray}
If $D = 3$, $\tilde{\tilde{U}}_{\mathrm{LL}}(i \omega_n, p)$ corresponds to the interaction that is mediated by a boson with linear dispersion,
\begin{eqnarray}
	&& \tilde{\tilde{U}}_{\mathrm{LL}}^{\mathrm{(3D)}}(i \omega_n, p) \propto \frac{1}{\omega_n^2 + (p v)^2} \, . \label{3Dboson}
\end{eqnarray}
If $D = 2$, $\tilde{\tilde{U}}_{\mathrm{LL}}(i \omega_n, p)$ can be envisioned as the interaction between 2D electrons that is mediated by a 3D collective soft mode with the propagator given by Eq.~(\ref{3Dboson}),
\begin{eqnarray}
	&& \hspace{-50pt}  \tilde{\tilde{U}}_{\mathrm{LL}}^{\mathrm{(2D)}}(i \omega_n, p_\parallel) \propto \int \frac{d p_{\mathrm{z}}}{2 \pi} \,  \tilde{\tilde{U}}_{\mathrm{LL}}^{\mathrm{(3D)}}(i \omega_n, p) \nonumber \\
	&& \hspace{10pt} \propto \frac{1}{\sqrt{\omega_n^2 + (p_\parallel v)^2}} \, , 
\end{eqnarray}
where $p_\parallel$ is the absolute value of 2D momentum, $\bm p = (\bm p_\parallel, p_{\mathrm{z}})$ is the 3D momentum of the collective mode.

We point out that the symmetry-breaking or Goldstone bosons such as acoustic phonons, do not provide interaction of the from of Eq.~(\ref{3Dboson}) due to the Adler's theorem.
Instead, the effective interaction in Eq.~(\ref{3Dboson}) may originate from soft collective modes near quantum criticality.
Such undamped critical bosons with linear dispersion are predicted in theories of continuous phase transitions in $d$-wave superconductors \cite{franz_mathrmqed_3_2002,franz_algebraic_2001}.

%\bibliographystyle{apsrev4-2}
%\bibliography{BOSO}
\end{document}